\documentclass[pre,preprintnumbers,amsmath,amssymb, aps ,superscriptaddress,twocolumn]{revtex4-1} 
\usepackage[utf8]{inputenc}
\usepackage{balance}
\usepackage[T1]{fontenc}
\usepackage{newtxtext}
\usepackage[varvw]{newtxmath}
\usepackage{graphicx}
\usepackage{color}
\usepackage{fixmath}
\usepackage{array}
\usepackage{bm}
\begin{document}

\title{Viscoelastic active diffusion governed by nonequilibrium fractional Langevin equations: underdamped dynamics and ergodicity breaking}

\author{Sungmin Joo}
\affiliation{Department of Physics, Pohang University of Science and Technology (POSTECH), Pohang 37673, Republic of Korea}

\author{Jae-Hyung Jeon}\email{jeonjh@gmail.com}
\affiliation{Department of Physics, Pohang University of Science and Technology (POSTECH), Pohang 37673, Republic of Korea}
\affiliation{Asia Pacific Center for Theoretical Physics (APCTP), Pohang 37673, Republic of Korea}


\begin{abstract} 
In this work, we investigate the active dynamics and ergodicity breaking of a nonequilibrium fractional Langevin equation (FLE) with a power-law memory kernel of the form $K(t)\sim t^{-(2-2H)}$, where $1/2<H<1$ represents the Hurst exponent. The system is subjected to two distinct noises: a thermal noise satisfying the fluctuation-dissipation theorem and an active noise characterized by an active Ornstein-Uhlenbeck process with a propulsion memory time $\tau_\mathrm{A}$.
We provide analytic solutions for the underdamped active fractional Langevin equation, performing both analytical and computational investigations of dynamic observables such as velocity autocorrelation, the two-time position correlation, ensemble- and time-averaged mean-squared displacements (MSDs), and ergodicity-breaking parameters. 
Our results reveal that 
the interplay between the active noise and long-time viscoelastic memory effect leads to unusual and complex nonequilibrium dynamics in the active FLE systems.  
Furthermore, the active FLE displays a new type of discrepancy between ensemble- and time-averaged observables. 
The active component of the system exhibits ultraweak ergodicity breaking where both ensemble- and time-averaged MSDs have the same functional form with unequal amplitudes. However, the combined dynamics of the active and thermal components of the active FLE system are eventually ergodic in the infinite-time limit. Intriguingly, the system has a long-standing ergodicity-breaking state before recovering the ergodicity. This apparent ergodicity-breaking state becomes exceptionally long-lived as $H\to1$, making it difficult to observe ergodicity within practical measurement times. Our findings provide insight into related problems, such as the transport dynamics for self-propelled particles in crowded or polymeric media. 
\end{abstract}
\maketitle
%

%


\section{\label{sec:intro}Introduction}

\begin{figure*}
 \centering
  \includegraphics[width=0.75\textwidth]{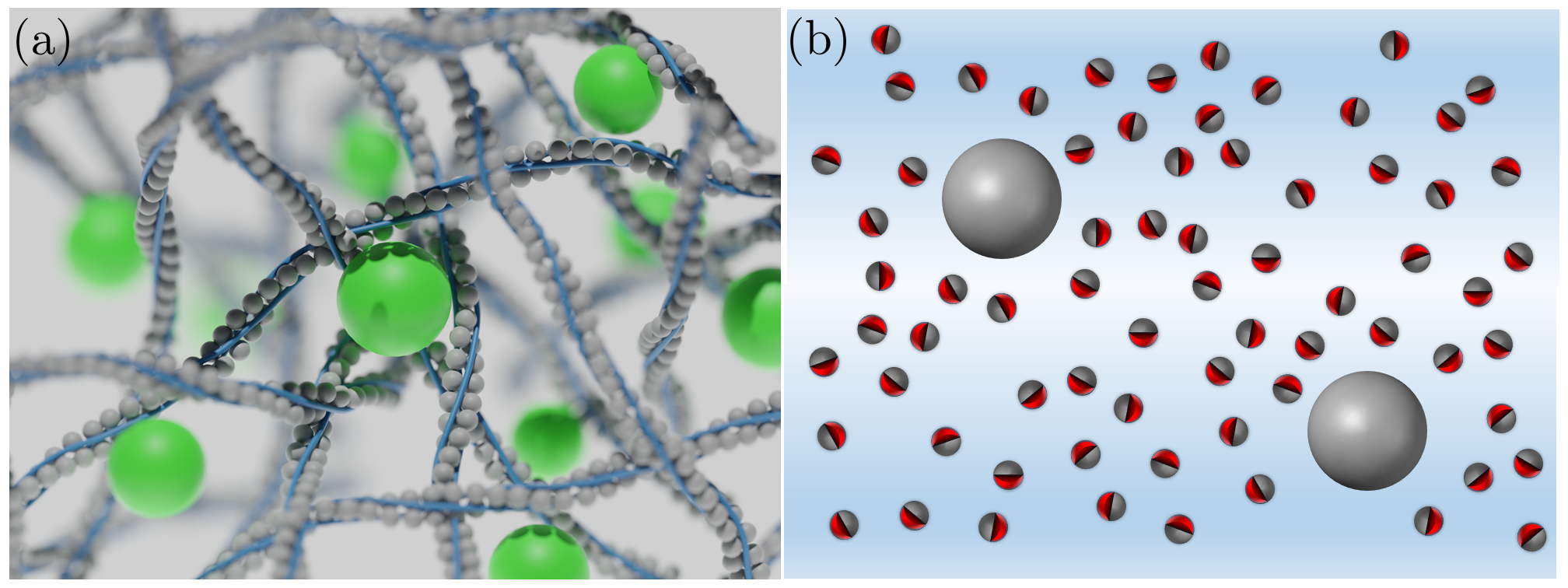}
 
\caption{A cartoon for active viscoelastic systems. (a) Active particles (green spheres) in a dense polymeric network represented by the gray spheres. The viscoelastic feedback originates from the interaction between the active tracers and background networks. (b) 
Passive tracer particles immersed in an active medium. The semi-red spheres depict self-propelled Janus particles. The passive tracer particles experience random active forces due to collisions with the self-propelled particles. These active particles collectively create a densely packed viscoelastic environment surrounding the tracers.
}
\label{fig1}
\end{figure*}

Active particles can be widely found in nature, appearing in various examples such as molecular motors-macromolecule complexes within living cell~\cite{Brangwynne2009,granick2015}, self-propelled Janus colloids~\cite{bocquet2010PRL,Rao2019Janusrod}, and biological swimmers ~\cite{Goldstein2012,Chamolly2017sperm}. Active particles exhibit self-propelled motion with directional and/or rotational memory, leading to dynamic behaviors that are far from equilibrium. 
Its nonequilibrium motions are modeled by several stochastic models, such as run-and-tumble model~\cite{srep2015Patteson,RevModPhys2016AP}, active Brownian particle~\cite{EPJ2012ABPRomanczuk,Shaebani2020ABP}, active Langevin particle~\cite{JCP2020Lowen}, and active Ornstein-Uhlenbeck particle (AOUP)~\cite{PRL2016AOUP,PRE2019Bonilla,AOUP2022Lowen}. The understanding the active diffusion has broader implications for active viscoelastic systems, where the nonequilibrium motion and viscoelastic effect of crowded environments coexist. Examples include the transport of passive tracer in active baths~\cite{Libchaber2000PRL} or in living cells~\cite{weihs2010PRE,Michael2000PRL}, chromosomal dynamics~\cite{PRL2009Bronstein,PNAS2022Trent}, lateral diffusion of membrane proteins~\cite{PRL2011Jeon,Lizana2016}, and the tracer diffusion in dense colloidal systems~\cite{Narinder2019,pnas2010ando}. Figure~\ref{fig1} illustrates a schematic representation of several active viscoelastic systems. Figure~\ref{fig1}(a) describes a typical model of intracellular environments, i.e., nonequilibrium biopolymer networks. Active particles (green sphere) diffuse in polymeric environments (gray sphere), such as acto-myosin network~\cite{PRL2004Weitzactin,Science2009Cooper}, endoplasmic reticulum network~\cite{LIN2014ER,PRE2018WeissER}, and microtubules~\cite{JCB1988microtubule,Michael2000PRL}, or ATP-consuming macromolecules strongly bound to polymer strands, such as DNA~\cite{Sanchez2012} and chromosomes~\cite{PRL2009Bronstein,Bronshtein2015naturecomm}. The presence of embedded polymers leads to viscoelastic feedback to active particles within the network. 
In Fig.~\ref{fig1}(b), thermal tracers (gray spheres) are placed in an active bath. Dense active particles (semi-red spheres) provide athermal noises to the tracers as well as lead to viscoelastic feedbacks~\cite{PRL2022Granek,PRLLubensky2003,Igor2002PNAS}. 

To describe the diffusion dynamics in these active complex systems, fractional Langevin equations (FLEs) have emerged as a valuable framework. These equations take the form:
\begin{equation}\label{eq:GLE}
    \int_0^t K(t-t')\dot{x}(t')dt'=\zeta_\mathrm{th}(t)+\zeta_\mathrm{ac}(t).
\end{equation}
Here, the memory kernel $K(t)$ accounts for a viscoelastic relaxation of a given embedding medium. Among a variety of viscoelastic systems, we here consider the systems that, at the timescales of interest where the many-body collective dynamics emerge, $K(t)$ is described by a power-law function
\begin{equation}
    K(t)\sim t^{-\alpha}
\end{equation}
where $0<\alpha<1$. For example, a single monomer in a flexible polymer is governed by $K(t)\sim t^{-1/2}$~\cite{PRE2010FLEEli,PRE2015Sakaue,Vandebroek2017JSP,SoftMatter2020Joo} and $K(t)\sim t^{-2\nu/(1+2\nu)}$ if the excluded volume interactions are present ($\nu$: Flory exponent)~\cite{DeGennes1976,Panja2010JSM,PRL2019Netz}. 
In a semi-flexible filament, the memory kernel behaves as $K(t)\sim t^{-3/4}$~\cite{granek1997EDP,Michael2000PRL,Han2023}. Moreover, $K(t)\sim t^{-2/3}$ for the height undulation dynamics of a planar membrane~\cite{granek1997EDP,Sung2018} as well as a single monomer dynamics in the Zimm model~\cite{Panja2010JSM,Sung2018}. In the presence of polymer condensation and entanglement, the memory kernel is given by $K(t)\sim t^{-1/4}$ for a single monomer in a reptating polymer~\cite{Gary2016JCPreptation}. 
The $\zeta_\mathrm{th}(t)$ is a thermal noise whose auto-covariance is related to the memory kernel via the fluctuation-dissipation theorem (FDT), i.e., $K(t-t')=k_B\mathcal{T}\langle\zeta_\mathrm{th}(t)\zeta_\mathrm{th}(t')\rangle$. The $\zeta_\mathrm{ac}(t)$ represents an active noise that violates FDT and usually has a finite memory time. It is responsible for the self-propelled movement of the active particle~\cite{EPJ2012ABPRomanczuk,buttinoni2013dynamical,Shaebani2020ABP} or mimics a nonequilibrium ambient noise from the active bath~\cite{Libchaber2000PRL,Santra2023activebath,Shaebani2020ABP}. The special case of $\zeta_\mathrm{ac}(t)=0$ is referred to as the equilibrium FLE. This equilibrium model has been employed to study the diffusion dynamics in thermal viscoelastic media~\cite{QIAN2000137,Chung2017,Softmatter2020Bonfanti,IOP2013Hofling} induced by macromolecular crowding and/or hydrodynamics interactions ~\cite{Kupferman2004,SandD2002Kupferman,JChemJ2019Lee,PhysicaA2015FODOR}. The resulting diffusion dynamics are typically viscoelastic subdiffusion where the mean-squared displacement (MSD) scales as 
\begin{equation}\label{eq:msdalpha}
\langle x^2(t)\rangle\sim t^{\mu}~\hbox{with}~\mu=\alpha.
\end{equation} 
Examples are the tracer diffusion in crowded dextran matrices~\cite{PRLDextran2009}, micellar solutions~\cite{Szymanski2006,Jeon2013Micellar}, and lipid membranes~\cite{PREFlenner2009lipid,PRL2012Jeonlipid,FD2013lipid}.

For the nonequilibrium case of the FLE~\eqref{eq:GLE} with $\zeta_\mathrm{ac}(t)\neq 0$, previous studies reported the following properties. When the active noise has an exponentially decaying correlation having $\langle \zeta_\mathrm{ac}(t)\zeta_\mathrm{ac}(0)\rangle\sim \exp(-t/\tau_\mathrm{A})$, the viscoelastic active diffusion exhibits $\langle x^2(t)\rangle\sim t^{2\alpha}$ for $t<\tau_\mathrm{A}$ and $\langle x^2(t)\rangle\sim t^{2\alpha-1}$ for $t>\tau_\mathrm{A}$~\cite{PRE1996Porra,SoftMatter2020Joo,Han2023,Michael2000PRL,PRE2005BaoOUN,Softmatter2017Vandebroek,Grimm2018Softmatter,Sakaue2017Softmatter}. For another type of active noises having a long-range correlation of a power-law form $\langle \zeta_\mathrm{ac}(t)\zeta_\mathrm{ac}(0)\rangle\sim t^{-\beta}$, the overdamped long-time dynamics follow the MSD of $\langle x^2(t)\rangle\sim t^{2\alpha-\beta}$~\cite{PRA1991Muralidhar,PRE1996Porra,Michael2002PRE,PRE2006Kwok,PRE2008Vinales,PRX2014Eli}. Compared to the equilibrium viscoelastic subdiffusion with $\alpha=\beta$, the active viscoelastic diffusion can exhibit more diverse dynamic patterns from superdiffusion, Fickian, subdiffusion, ultraslow diffusion ($\sim\ln t$) to confined motion.


Extending the FLE~\eqref{eq:GLE}, here we investigate the active viscoelastic diffusion governed by an underdamped nonequilibrium FLE: 
\begin{equation}\label{eq:FLE}
m\ddot{x}(t)+\int_0^t K(t-t')\dot{x}(t')dt'=\zeta_\mathrm{th}(t)+\zeta_\mathrm{ac}(t).
\end{equation}
In the absence of $\zeta_\mathrm{ac}(t)$, the above FLE is the equilibrium underdamped FLE~\cite{PRE1996Porra,PRE2013Ralf,PRE2021Ralf,Bao2017,PRE2008Burov}. This equilibrium model undergoes ballistic dynamics $\langle x^2(t)\rangle\sim t^2$ regardless of $\alpha$ (when $v_0\neq 0$) in the underdamped regime and cross-overs to subdiffusion with $\langle x^2(t)\rangle\sim t^{\alpha}$ in the overdamped regime~\cite{Frick2009SIAM,JRheol2009Gregory,PRE2014GLEunder}. 
As an exceptional case, when $v_0=0$, the system exhibits hyperdiffusion with $\langle x^2(t)\rangle\sim t^{4-\alpha}$ in the underdamped regime. 
It was demonstrated that micron-sized particles in an actin filament network and membrane proteins in a lipid bilayer perform the viscoelastic thermal diffusion from the ballistic to a subdiffusion~\cite{PRE2013Sylvia,DiCairano2021protein}. 

In the case of active FLE systems ($\zeta_\mathrm{ac}(t)\neq0$), little is known about its full-time dynamics and ergodic properties. Here, we analytically and numerically solve the active FLE~\eqref{eq:FLE} and systematically investigate the dynamic properties of this non-equilibrium process. 
It turns out that compared to the equilibrium counterpart the active FLE system reveals unexpectedly intricate dynamic patterns as consequences of the combined effects of the non-Markovian active noise, the underdamped inertia dynamics, and the long-time viscoelastic feedback. 
In contrast to the equilibrium FLE, the active FLE exhibits multi-scaling underdamped dynamics involving a combination of ballistic and hyperdiffusion ($\sim t^4$, $\sim t^3$). After the underdamped dynamics, the system undergoes one of the two distinct overdamped dynamics, depending on the relative magnitude of the momentum relaxation time and the noise memory time ($\tau_\mathrm{A}$). If the former is larger than the latter, the overdamped dynamics is simply the active subdiffusion ($\sim t^{2\alpha-1}$) followed by the thermal subdiffusion, Eq.~\eqref{eq:msdalpha}, beyond a cross-over time $\tau^*$. For the other case, the system has the three-scaling overdamped dynamics consisting of the active superdiffusion ($\sim t^{2\alpha}$), active subdiffusion ($\sim t^{2\alpha-1}$), and finally thermal subdiffusion ($\sim t^{\alpha}$). 

In this study, additionally, we are interested in time-averaged observables and the ergodicity of the active FLE system. To our best knowledge, our study is the first analytic work to evaluate the time-averaged MSD and the ergodicity-breaking parameter. It turns out that the time-averaged MSDs significantly differ from the ensemble-averaged MSDs. In the underdamped regime, the activity-induced hyperdiffusion is absent. Instead, it simply manifests the ballistic dynamics. In the overdamped regime, the active dynamics exhibit \textit{ultraweak} ergodicity breaking in which both ensemble- and time-averaged MSDs have the same scaling function but differ in their amplitudes. We also find that the activity-induced ergodicity-breaking phenomena are apparently long-standing before the active FLE system finally enters the ergodic regime for $t\gg\tau^*$. The cross-over time $\tau^*$ has a non-monotonic dependence on the memory kernel exponent $\alpha$. Importantly, $\tau^*$ rapidly increases as $\alpha$ approaches zero. Within practical observation times, accordingly, the active FLE system cannot reach the ergodic state, residing in the \textit{apparent} ergodicity-breaking state.

The paper is organized in the following. In Sec.~\ref{sec2}, we provide a comprehensive explanation of our active FLE model, including the exact solutions for velocity and position variables. Additionally, we encapsulate the simulation method to solve the active FLE. In Sec.~\ref{sec3}, we present the main results of our work. This encompasses analytic and simulation studies of velocity autocorrelation, ensemble- and time-averaged MSDs, as well as an examination of the ergodicity-breaking parameters. 
Lastly, Sec.~\ref{sec4} summarizes our primary findings while discussing the practical application of our theories in diverse active viscoelastic systems.


\section{The Model \& Method} \label{sec2}

\subsection{The active fractional Langevin equation model}

Within the framework of the generalized Langevin equation~\eqref{eq:FLE}, we introduce a class of nonequilibrium dynamic models referred to as the active fractional Langevin equation (FLE).
This model incorporates a viscoelastic memory effect with a power-law kernel $K(t)$ in the following form:
\begin{equation}\label{eq:FLEwoFDT2}
     m\ddot{x}(t)+\gamma_0\int_0^t(t-t')^{2H-2}\dot{x}(t')dt'=\eta\xi_{H}(t)+\eta_\mathrm{A}\xi_\mathrm{A}(t).
\end{equation} 
Here, the Hurst exponent $H$ is restricted to the range of $1/2<H<1$ such that the memory term can be expressed as
\begin{equation}
    \gamma_0\int_0^t(t-t')^{2H-2}\dot{x}(t')dt'=\gamma_0\Gamma(2H-1)\frac{d^{2-2H}}{d t^{2-2H}}x(t).
\end{equation}
In this expression, $\frac{d^{2-2H}}{d t^{2-2H}}f(t)=\Gamma(2H-1)^{-1}\int_0^t (t-t')^{2H-2} \frac{d}{dt'}[f(t')]dt'$ represents the Caputo fractional derivative of order $2-2H$ (with $0<2-2H<1$)~\cite{caputo2014}. 
The $\gamma_0(>0)$ is a generalized frictional coefficient with the physical dimension of [$\mathrm{kg}\cdot\mathrm{s}^{-2H}$]. In RHS of Eq.~\eqref{eq:FLEwoFDT2}, the first term describes the thermal equilibrium noise, which is connected to the memory kernel via FDT, thus, 
\begin{equation}\label{FDT}
\eta^2\langle \xi_H(t)\xi_H(t')\rangle=\gamma_0 k_B\mathcal{T} |t-t'|^{2H-2}.
\end{equation}
The thermal noise, which follows the power-law auto-covariance given by Eq.~\eqref{FDT}, is realized by fractional Gaussian noise (fGn)~\cite{SIAMreview1968}. fGn is a stationary incremental Gaussian process with zero mean $\langle\xi_H(t)\rangle=0$ and auto-covariance $\langle\xi_H(t)\xi_H(t')\rangle=2H(2H-1)|t-t'|^{2H-2}$ with $H\neq1/2$.
The coupling constant $\eta$ is then determined to be $\eta=\sqrt{\frac{k_B\mathcal{T}\gamma_0}{2H(2H-1)}}$. The FLE~\eqref{eq:FLEwoFDT2} without the active noise $\xi_\mathrm{A}$ is commonly referred to as the equilibrium FLE, which describes viscoelastic subdiffusion under equilibrium conditions~\cite{PRE2001Lutz,PRE2008Burov,PRE2010Jeon,PRE2013Ralf,PRE2013Sylvia,JRheol2009Gregory,supercooledwater2023PCCP,PRLDextran2009}.

The $\xi_\mathrm{A}$ with the amplitude $\eta_\mathrm{A}=\sqrt{k_B\mathcal{T}\gamma_0}$ in Eq.~\eqref{eq:FLEwoFDT2} represents the active noise responsible for the self-propulsion motion of an active particle or a nonequilibrium noise from an active heat bath. Following Refs.~\cite{Libchaber2000PRL,PRE2019Bonilla,SoftMatter2020Joo,macromolecules2022YJKim,PRE2023Santra}, we model the active noise with a symmetric Gaussian process with a finite memory time described by an exponentially decaying auto-covariance
\begin{equation}
    \langle\xi_\mathrm{A}(t)\xi_\mathrm{A}(t')\rangle= D_\mathrm{A}e^{-\frac{|t-t'|}{\tau_\mathrm{A}}}.
\end{equation}
Here, $\tau_\mathrm{A}$ is the propulsion memory time and $D_\mathrm{A}$ is the (dimensionless) noise strength. Such Gaussian active noises are known as active Ornstein-Uhlenbeck (OU) noise, which is governed by the following Langevin equation~\cite{OUprocess1930PR}
\begin{equation}\label{eq:genaoup2}
\frac{d\xi_\mathrm{A}(t)}{dt}=-\frac{1}{\tau_\mathrm{A}}\xi_\mathrm{A}(t)+\sqrt{\frac{2D_\mathrm{A}}{\tau_\mathrm{A}}}\varkappa(t)
\end{equation} 
where $\varkappa(t)$ is a delta-correlated Gaussian noise of $\langle\varkappa(t)\rangle=0$ and $\langle\varkappa(t)\varkappa(t')\rangle=\delta(t-t')$.
The active particle driven by the active OU noise is referred to as active Ornstein-Uhlenbeck particle (AOUP)~\cite{PRL2016AOUP,PRE2019Bonilla,AOUP2022Lowen}. In this sense, the active FLE~\eqref{eq:FLEwoFDT2} describes the nonequilibrium diffusion dynamics of AOUPs embedded in a viscoelastic environment. 

It is worthwhile to introduce two limiting models of the active FLE~\eqref{eq:FLEwoFDT2}. When $H\to1/2$, it becomes the ordinary underdamped AOUP model~\cite{AOUP2022Lowen}
\begin{equation}\label{AOUPmodel}
     m\ddot{x}(t)+\gamma_0\dot{x}(t)=\eta\xi_{H}(t)+\eta_\mathrm{A}\xi_\mathrm{A}(t).
\end{equation} 
This model explains the diffusion dynamics of AOUPs in a \textit{viscous} fluid~\cite{PRL2016AOUP,PRE2019Bonilla}. When $H\to1$, the memory kernel term reduces to $\gamma_0[x(t)-x(0)]$, so the corresponding Langevin equation is responsible for AOUPs in an elastic material~\cite{PREAOUPharmonic}. For $H\in(1/2,~1)$, the system is viscoelastic.

The active FLE~\eqref{eq:FLEwoFDT2} can be analytically solved in the Laplace space ($\tilde{f}(s)\equiv\mathcal{L}[f(t)]=\int_0^\infty e^{-st}f(t)dt$). Considering the initial conditions $\dot{x}(0)=v_0$ 
and $x(0)=0$, the Laplace-transformed velocity is given by
\begin{equation}\label{eq:vtilde2}
     \tilde{v}(s)=\frac{ v_0+\eta m^{-1} \widetilde{\xi_{H}}(s)+\eta_\mathrm{A}m^{-1}\widetilde{\xi_\mathrm{A}}(s)}{s+\tau_H^{-2H}s^{1-2H}}
\end{equation}
where $\tau_H=(m/[\gamma_0\Gamma(2H-1)])^{1/2H}$ is a characteristic time of the system. Using the generalized Mittag-Leffler function~\cite{MLFunction1955,MLfunction2011} 
\begin{equation}\label{eq:MLF}
    E_{a,b}(z)=\sum_{k=0}^\infty\frac{z^k}{\Gamma(a k+b)}
\end{equation} 
where $a,b\in \mathbb{R}$ and $z\in \mathbb{C}$ (see Appendix~\ref{sec:appendixB} for further information),
the analytic expression for the velocity $v(t)$ is obtained as follows:
\begin{equation}\label{eq:velocity2}
    \begin{aligned}
        v(t)=&v_0E_{2H,1}\left[-\left(\frac{t}{\tau_H}\right)^{2H}\right]\\
    &+\frac{\eta}{m}\int_0^t\xi_{H}(t')E_{2H,1}\left[-\left(\frac{t-t'}{\tau_H}\right)^{2H}\right]dt'\\
    &+\frac{\eta_\mathrm{A}}{m}\int_0^t\xi_\mathrm{A}(t')E_{2H,1}\left[-\left(\frac{t-t'}{\tau_H}\right)^{2H}\right]dt'.
    \end{aligned}
\end{equation}
In the limit of $H\to1/2$ (viscous fluids), the Mittag-Leffler function is simplified to $E_{1,1}(x)=\exp(-x)$ and Eq.~\eqref{eq:velocity2} is the solution to the ordinary Langevin equation~\eqref{AOUPmodel}. In this case, the system's relaxation dynamics is described by an exponential function with the momentum relaxation time $\tau_{1/2}=m/\gamma_0$. For $1/2<H<1$, the relaxation function $E_{2H,1}[-(t/\tau_H)^{2H}]\approx \exp[-(t/\tau_H)^{2H}/\Gamma(2H+1)]$ for $t\ll\tau_H$ and $\approx \frac{\tau_H^{2H}}{\Gamma(1-2H)}t^{-{2H}}$ for $t\gg\tau_H$. Accordingly, the $\tau_H$ can be viewed as the momentum relaxation time for the underdamped dynamics of a viscoelastic system to relax.

The velocity, Eq.~\eqref{eq:velocity2}, can be viewed as the sum of the two contributions: 
\begin{equation}
v(t)=v_\mathrm{th}(t)+v_\mathrm{ac}(t).    
\end{equation}
The former is the solution for the equilibrium FLE without the active noise~\cite{PRE2013Ralf,PRE2021Ralf,Bao2017}, i.e., 
\begin{equation}\label{eq:v_th} 
\begin{aligned}
v_\mathrm{th}(t)=&v_0E_{2H,1}\left[-\left(\frac{t}{\tau_H}\right)^{2H}\right]\\
&+\frac{\eta}{m}\int_0^t\xi_{H}(t')E_{2H,1}\left[-\left(\frac{t-t'}{\tau_H}\right)^{2H}\right]dt'.
\end{aligned}
\end{equation}
The latter is the contribution of the active noise
\begin{equation}\label{eq:v_ac}
    v_\mathrm{ac}(t)=\frac{\eta_\mathrm{A}}{m}\int_0^t\xi_\mathrm{A}(t')E_{2H,1}\left[
-\left(\frac{t-t'}{\tau_H}\right)^{2H}\right]dt'.
\end{equation}

The position $x(t)$ can be obtained analytically using the similar Laplace transform technique. The position is also written as the sum of the thermal and active parts, i.e., $x(t)=x_\mathrm{th}(t)+x_\mathrm{ac}(t)$ where
\begin{equation}\label{eq:xeq}   
\begin{aligned}
    x_\mathrm{th}(t)=&v_0 tE_{2H,2}\left[-\left(\frac{t}{\tau_H}\right)^{2H}\right]\\
    &+\frac{\eta}{m}\int_0^t\xi_{H}(t')t'E_{2H,2}\left[-\left(\frac{t-t'}{\tau_H}\right)^{2H}\right]dt'
\end{aligned}
\end{equation}
is the solution for the equilibrium FLE~\cite{PRE2013Ralf,PRE2021Ralf,Bao2017} and 
\begin{equation}\label{eq:xac}
    x_\mathrm{ac}(t)=\frac{\eta_\mathrm{A}}{m}\int_0^t\xi_\mathrm{A}(t')t'E_{2H,2}\left[-\left(\frac{t-t'}{\tau_H}\right)^{2H}\right]dt'.
\end{equation}


\subsection{Numerical method}
To complement our analytic study, we numerically solve our FLE model for various parameter conditions and perform the Langevin dynamics simulation. The numerical technique to solve the active FLE is explained in detail in Appendix~\ref{sec:appendixA}. Here we briefly outline the simulation scheme. The same protocol is used in Refs.~\cite{PRE2009Eli,PRE2010Jeon,Numerics2013Guo}. In simulating the active FLE~\eqref{eq:FLEwoFDT2}, we set $\gamma_0=1$ or $100$, $m=1$, $k_B\mathcal{T}=1$, and $H\in\{5/8,~3/4,~7/8\}$. 
The propulsion time of the active OU noise was set to be $\tau_\mathrm{A}=0.01$, $0.1$, $1$, and $10$. The noise strength was $D_\mathrm{A}=0$ (thermal), $25$, $100$, and $400$. 
In the simulation, the initial position was always fixed to $x_0=0$. On the other hand, we studied the $ v_0$-dependence on the transport dynamics: For the case of a fixed initial velocity, it was put to be $v_0=0$. For a Gaussian-distributed random initial velocity, it was chosen to have the variance $\langle v_0^2\rangle_\mathrm{th}=k_B\mathcal{T}/m$ for the FDT-satisfying equilibrium systems and $\langle v_0^2\rangle=\langle \overline{v^2}\rangle_\mathrm{st}$ (i.e., the stationary value) for the FDT-violating systems. 

We conducted simulations with a total observation time of $T=Nh$, where the total time step was set to $N=15\times 10^3$--$10^8$ and the integration time was $h=10^{-4}$, $10^{-3}$, or $10^{-2}$. Each parameter set was simulated using a range of $500$--$20000$ runs to ensure statistically significant ensemble-averaged results. Prior to the main numerical simulations, we rigorously validated our simulation code by applying it to solve equilibrium FLE systems reported in Ref.~\cite{PRE2013Ralf} for various initial conditions. The simulation results were in excellent agreement with the existing analytical theory. Further details about this validation process can be found in Appendix~\ref{sec:appendixA} \& Fig.~\ref{figA1}.

\section{Results: Active diffusion in viscoelastic media}\label{sec3}

\subsection{Velocity autocorrelation function (VACF) }\label{sec:VACF}

\begin{figure*}
 
      \includegraphics[width=0.75\textwidth]{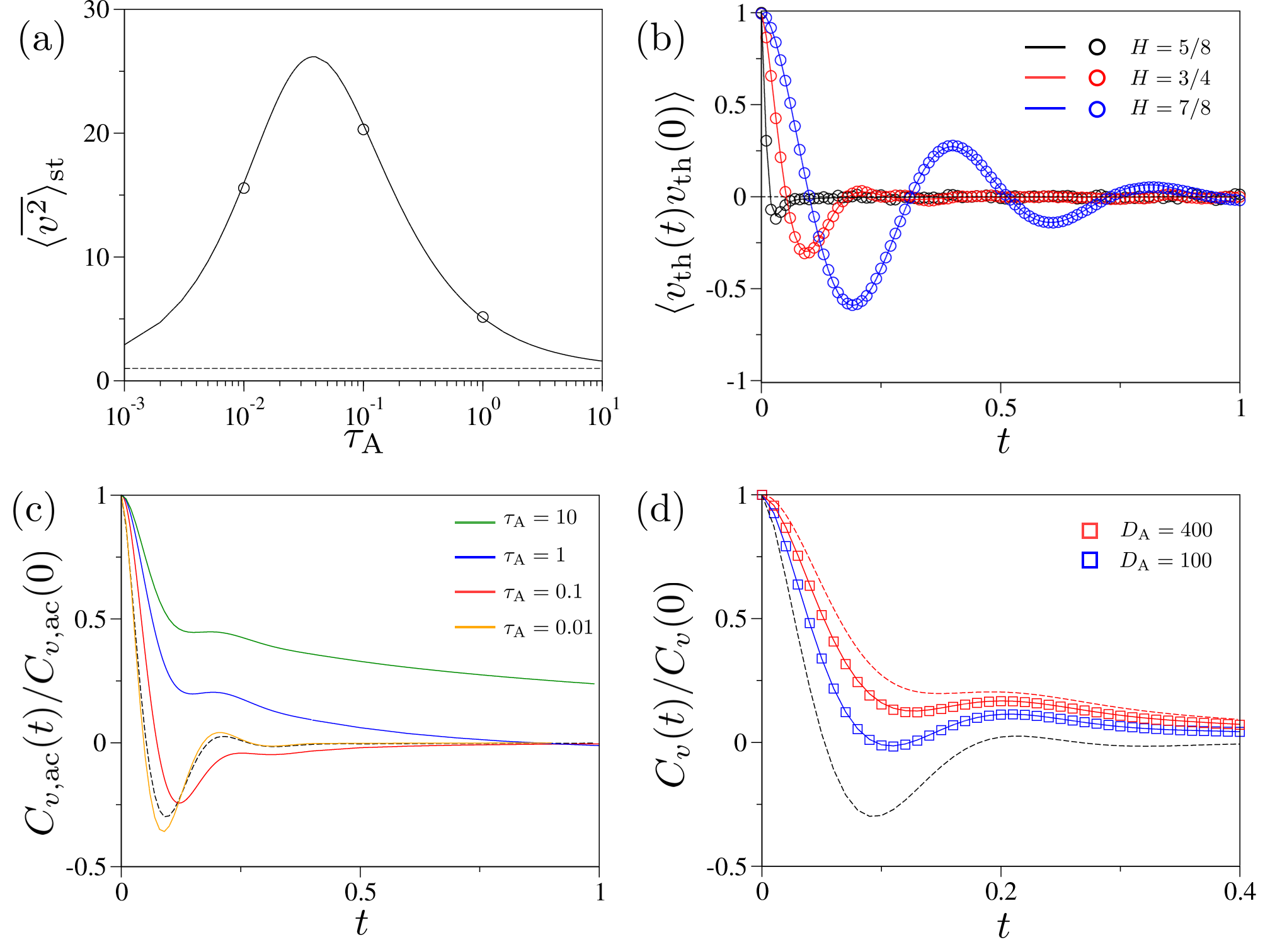}
 
\caption{Simulation results for the velocity variable with theoretical prediction. (a) The mean-squared velocity, $\langle \overline{v^2}\rangle_\mathrm{st}$, as a function of $\tau_\mathrm{A}$ at $H=3/4$. $\langle \overline{v^2}\rangle_\mathrm{st}$ has a maximum near $\tau_H\approx0.03$. The simulation data (circles) are successfully explained by Eq.~\eqref{eq:vsquare_active} (solid line). The dashed line represents $\langle \overline{v^2}\rangle_\mathrm{st}=\langle v^2\rangle_\mathrm{th}(=1)$. (b--d) The VACFs with theoretical expectation with various conditions for $H$, $\tau_\mathrm{A}$, and $D_\mathrm{A}$. 
(b) Simulation data for $\langle v_\mathrm{th}(t)v_\mathrm{th}(0)\rangle$ with $H=5/8$, $3/4$, and $7/8$. The ensemble-averaged VACFs are obtained from $20000$ trajectories for given parameters. The theoretical lines [Eq.~\eqref{eq:VACFth}] are numerically plotted, and they show good agreement with the simulation results (symbols).
(c) $C_{v,\mathrm{ac}}(t)/C_{v,\mathrm{ac}}(0)$ [Eq.~\eqref{eq:VACFac}] is plotted for $\tau_\mathrm{A}=0.01$, $0.1$, $1$, $10$, and $H=3/4$. For comparison, the thermal VACF of $C_{v,\mathrm{th}}(t)/C_{v,\mathrm{th}}(0)$ is plotted with the black dashed line. 
(d) $C_{v}(t)/C_{v}(0)$ with $D_\mathrm{A}=100$, $400$, $\tau_\mathrm{A}=1$, and $H=3/4$. The black dashed line represents the thermal VACF [Eq.~\eqref{eq:VACFth}] at $D_\mathrm{A}=0$ while the red dashed line represents Eq.~\eqref{eq:VACFac}. The solid lines show our theory [Eq.~\eqref{eq:Cdeltaactive}] and good agreement with our simulation results (symbols).}
\label{fgr2}
\end{figure*}

The two-time VACF $\langle v(t)v(t')\rangle$ can be formally obtained using the solution $v(t)$ [Eq.~\eqref{eq:velocity2}]. Based on the fact that the thermal and active noises are independent of each other, we can write the VACF as the sum of the thermal and active VACFs: $\langle v(t)v(t')\rangle=\langle v_\mathrm{th}(t)v_\mathrm{th}(t')\rangle+\langle v_\mathrm{ac}(t)v_\mathrm{ac}(t')\rangle$. The thermal VACF is evaluated in accordance with Ref.~\cite{Bao2017} as:
\begin{equation}
\label{eq:vacfth}
\langle v_\mathrm{th}(t)v_\mathrm{th}(t')\rangle=\langle v^2\rangle_\mathrm{th} F_1(\tau)+\left(\langle v_0^2\rangle-\langle v^2\rangle_\mathrm{th} \right) F_1(t)F_1(t'),
\end{equation}
where $\tau=t'-t$ is the time lag between two observation times ($t<t'$), $\langle v^2\rangle_\mathrm{th}=k_B\mathcal{T}/m$, and $F_1(t)=E_{2H,1}\left[-(t/\tau_H)^{2H}\right]$. The VACF for the thermal velocity $v_\mathrm{th}(t)$ satisfies the stationary condition at thermal equilibrium. The product term of $F_1(t)$ and $F_1(t')$ is ignored at the equilibrium initial condition of $\langle v_0^2\rangle=\langle v^2\rangle_\mathrm{th}$ or in the infinite-time limit of $t,t'\to \infty$. From Eq.~\eqref{eq:v_ac}, the active part of the VACF can be expressed as
\begin{equation}\label{eq:vacfac}
    \langle v_\mathrm{ac}(t)v_\mathrm{ac}(t')\rangle
= \frac{\eta_\mathrm{A}^2D_\mathrm{A}}{m^2}\int_0^{t} \{F_1(x+\tau)F_2(x)+ F_1(x)F_2(x+\tau)\}dx
\end{equation}
where 
$F_2(t)=\mathcal{L}^{-1}[\tau_\mathrm{A}(1+s\tau_\mathrm{A})^{-1}(s+\tau_H^{-2H} s^{1-2H})^{-1}](t)$. 
While $v_\mathrm{ac}(t)$ does not reach the equilibrium state, its VACF can satisfy the time-translation symmetry $\langle v_\mathrm{ac}(t)v_\mathrm{ac}(t')\rangle=\langle v_\mathrm{ac}(\tau)v_\mathrm{ac}(0)\rangle$ in the limit of $t,t'\to\infty$ due to the stationary nature of the active OU noise. Thus, we can introduce the stationary VACF as follows: 
\begin{equation}\label{eq:Cdeltaactive}
C_v(t)\equiv\lim_{t_0 \to\infty}\langle v(t_0+t)v(t_0)\rangle=C_{v,\mathrm{th}}(t)+C_{v,\mathrm{ac}}(t) 
\end{equation}
where 
\begin{equation}\label{eq:VACFth}
C_{v,\mathrm{th}}(t)=\langle v^2\rangle_\mathrm{th} E_{2H,1}\left[-\left(\frac{t}{\tau_H}\right)^{2H}\right]
\end{equation} 
and 
\begin{equation}\label{eq:VACFac}
C_{v,\mathrm{ac}}(t)=\frac{\eta_\mathrm{A}^2D_\mathrm{A}}{m^2}\int_0^{\infty }\{F_1(t'+t)F_2(t')dt'+F_1(t')F_2(t'+t)\}dt'.
\end{equation} 
In these expressions, $t$ acts as time lag. Now let us examine the behaviors of $C_v(t)$ upon varying the active noise. 

First, the stationary mean-squared velocity, i.e., the kinetic energy of the active FLE system, is obtained by evaluating $C_v(t\to0)$, yielding
\begin{equation}\label{eq:vsquare_active}
    \langle \overline{v^2}\rangle_\mathrm{st}=\langle v^2\rangle_\mathrm{th} \left(1+2\frac{\gamma_0D_\mathrm{A}}{m}\int_0^{\infty} F_1(t')F_2(t')dt'\right).
\end{equation}
Here, the second term explains the contribution of the active component of the velocity. This term dominates over the thermal counterpart when the active noise is sufficiently stronger than the thermal noise. The mathematical expression suggests that the mean-squared velocity monotonically increases with the strength of the active noise $D_\mathrm{A}$. On the other hand, it has a non-monotonic dependence on $\tau_\mathrm{A}$. Figure~\ref{fgr2}(a) shows $\langle \overline{v^2}\rangle_\mathrm{st}$ as a function of $\tau_\mathrm{A}$ where the simulation result (symbol) is plotted together with the theoretical expression~\eqref{eq:vsquare_active} (solid line). 
Notably, $\langle \overline{v^2}\rangle_\mathrm{st}$ has a uni-modal profile with the maximum occurring at $\tau_\mathrm{A}\simeq\tau_H~(\approx0.03)$. 
This indicates that the energy injection into the system through the active OU noise reaches its maximal efficiency when $\tau_\mathrm{A}\approx \tau_H$. This behavior can be understood as follows. In the limit of 
$\tau_\mathrm{A}\approx0$, the active noise changes too rapidly compared to the viscoelastic relaxation, so the active noise hardly injects energy into the system; in this limit, $\langle \overline{v^2}\rangle_\mathrm{st}=\langle v^2\rangle_\mathrm{th} \left(1+\frac{\gamma_0D_\mathrm{A}\tau_\mathrm{A}}{m} \int_0^\infty F_1(t')^2dt'\right)\approx\langle v^2\rangle_\mathrm{th}(1+\mathcal{O}(\tau_\mathrm{A}))$. 
In the regime where $0\ll\tau_\mathrm{A}<\tau_H$, a longer propulsion memory time allows the system to move coherently with the viscoelastic directional movement, resulting in an increase in kinetic energy. However, in the opposite regime where $\tau_\mathrm{A}>\tau_H$, the active noise hampers the system's directional motion, leading to the dissipation of the active energy into the environment. The energy dissipation becomes more significant as $\tau_\mathrm{A}$ increases. In the limit of $\tau_\mathrm{A}\to\infty$ (i.e., the constant-force limit), it is found that 
$\int_0^{\infty} F_1(t')F_2(t')dt'= F_2(t)^2|^{\infty}_0=0$, and thus $\langle \overline{v^2}\rangle_\mathrm{st}$ converges to $\langle v^2\rangle_\mathrm{th}$.

In Fig.~\ref{fgr2}(b) we plot the profile of $C_{v,\mathrm{th}}(t)$ for several values of $H$ (symbols: simulation, lines: Eq.~\eqref{eq:VACFth}). The thermal component of the VACF generally shows a negative correlation. As the system has a more elastic component ($H\to1$), the negative correlation becomes stronger, and the thermal VACF exhibits oscillatory behaviors.
Our numerical analysis reveals that the oscillation starts at $H\approx 0.7111$, and these oscillations become more pronounced as $H$ approaches 1.

In Fig.~\ref{fgr2}(c) we examine the behaviors of the active VACF, $C_{v,\mathrm{ac}}(t)$, for increasing $\tau_\mathrm{A}$ at fixed $H$. It exhibits distinct characteristics depending on $\tau_\mathrm{A}$. 
When $\tau_\mathrm{A}\ll\tau_H$, $C_{v,\mathrm{ac}}(t)$ has a very similar profile with $C_{v,\mathrm{th}}(t)$. For example, compare the active VACF for $\tau_\mathrm{A}=0.01$ with $C_{v,\mathrm{th}}(t)$ (plotted in the dashed line). Both curves are almost the same over the entire time window. Such agreement is consistently observed for other values of $H$ (not shown). Given the fact that $C_{v,\mathrm{th}}(t)\sim F_1(t)$ is the viscoelastic relaxation function of the medium (see Eqs.~\eqref{eq:v_th} and~\eqref{eq:v_ac}), driving the system with a delta-correlated-like athermal noise is expected to give dynamic information on the medium's inherent viscoelastic response 
\footnote{Rigorously, in the limit of $\tau_\mathrm{A}\to0$ for a fixed $D_\mathrm{A}$, $C_\mathrm{v,ac}(t)
\approx2A\int_0^{\infty} F_1(x+t)F_1(x)dx$ while $C_{v,\mathrm{th}}(t)\sim F_1(t)$. Although $C_\mathrm{v,ac}(t)$ does not precisely converges to $C_{v,\mathrm{th}}(t)$ as $\tau_\mathrm{A}\to0$, two functions are numerically very similar to each other.}. 
Evidently, $C_{v,\mathrm{ac}}(t)$ deviates from the thermal VACF with increasing $\tau_\mathrm{A}$. It has a pronounced positive correlation regime for large values of $\tau_\mathrm{A}$, which lasts roughly up to $t\approx\tau_\mathrm{A}$. 
In this regime, the system is driven by a nearly constant active force, causing its velocity to be in the same direction as the force.
In the other regime of $t\gg \tau_\mathrm{A}$, the active VACF eventually becomes negative due to the effect of the viscoelastic medium, decaying as $C_{v,\mathrm{ac}}(t) \sim t^{1-4H}$ irrespective of $\tau_\mathrm{A}$.

The profile of $C_v(t)$ is given by the sum of the thermal and active VACFs. Figure~\ref{fgr2}(d) shows the simulated $C_v(t)$ with the theoretical expression~\eqref{eq:Cdeltaactive} for $D_\mathrm{A}=100$ and $400$. For comparison, we also plot the theoretical lines for $C_{v,\mathrm{th}}(t)$ (dashed, black) and $C_{v,\mathrm{ac}}(t)$ (dashed, red). When the active noise is weak, it captures the essential features of $C_{v,\mathrm{th}}(t)$; it exhibits a negative correlation and oscillatory behaviors for large values of $H$. As the active noise is stronger, $C_{v}(t)$ tends to follow $C_{v,\mathrm{ac}}(t)$. Instead of the negative correlation, it has a positive correlation for $t<\tau_H$ and a weak negative correlation for $t>\tau_H$. 

\subsection{Mean-squared displacements (MSDs)}\label{sec:MSD}

\begin{figure*}
         \includegraphics[width=0.75\textwidth]{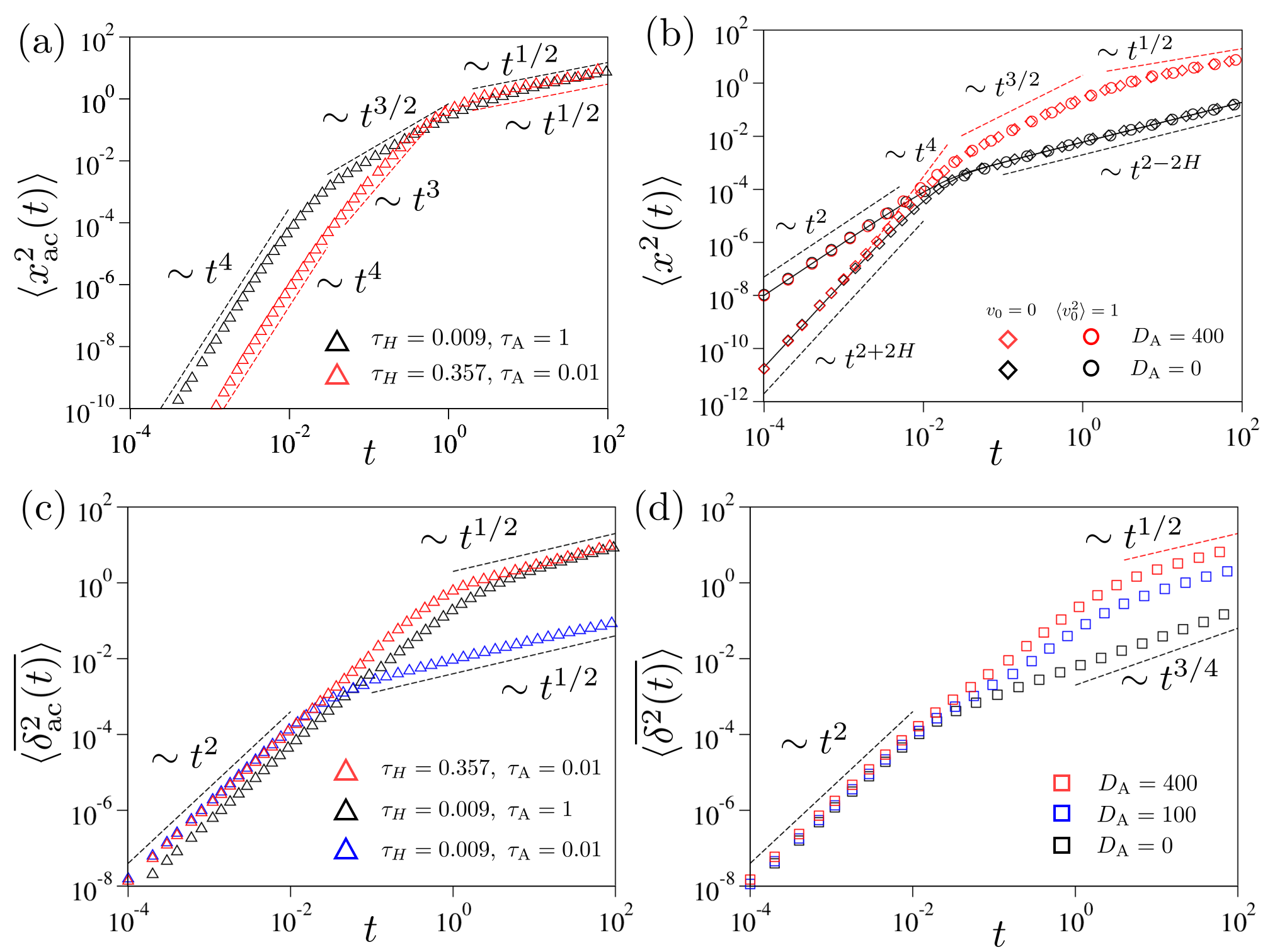}  
\caption{The ensemble- and time-averaged MSDs for $H=5/8$. 
(a) The comparison of the full-time dynamics of $\langle x_\mathrm{ac}^2(t)\rangle$ for Case $1$ ($\tau_\mathrm{A}>\tau_H$, black) \& Case $2$ ($\tau_H>\tau_\mathrm{A}$, red). Case $1$: $\tau_\mathrm{A}=1$, $\tau_H=0.009$. Case $2$: $\tau_\mathrm{A}=0.01$, $\tau_H=0.357$. The dashed lines depict the expected scaling from Eqs.~\eqref{eq:xac2_case1} \&~\eqref{eq:xac2_case2}. 
(b) The full-time dynamics of $\langle x^2(t)\rangle$ for $D_\mathrm{A}=0$ and $400$ with two initial conditions; $v_0=0$ and $\langle v_0^2\rangle=1$. Here the MSD is the result for Case 1 with $\tau_\mathrm{A}=1$ and $\tau_H=0.009$. In the plot, the black dashed lines represent the scalings of $t^2$, $t^{2+2H}$, and $t^{2-2H}$ expected in $\langle x_\mathrm{th}^2(t)\rangle$. The red dashed lines indicate the scalings of $t^4$, $t^{3/2}$, and $t^{1/2}$ expected in $\langle x_\mathrm{ac}^2(t)\rangle$. 
(c) The comparison of the time evolution of $\langle \overline{\delta_\mathrm{ac}^2(t)}\rangle$ for Case $1$ (black) and Case $2$ (red). Case $1$: $\tau_\mathrm{A}=1$, $\tau_H=0.009$. Case $2$: $\tau_\mathrm{A}=0.01$, $\tau_H=0.357$. As a special case, the result for $\tau_\mathrm{A}\approx \tau_H$ (blue) is presented with the parameters; $\tau_\mathrm{A}=0.01$ and $\tau_H=0.009$. 
(d) The profiles of $\langle \overline{\delta^2(t)}\rangle$ for varying $D_\mathrm{A}$. In this plot, Case 1 is plotted with $\tau_\mathrm{A}=1$ and $\tau_H=0.009$.}
\label{fgr3}
\end{figure*}
Following the same line of procedure in the previous section, we first evaluate the two-time position correlator $\langle x(t)x(t')\rangle$. 
According to Eqs.~\eqref{eq:xeq} and~\eqref{eq:xac}, $\langle x(t)x(t')\rangle$ is the combination of both the thermal and active contributions, which can be expressed as follows:
\begin{equation}\label{eq:xccth} 
\begin{aligned}
    \langle x_\mathrm{th}(t)x_\mathrm{th}(t')\rangle=&(\langle v_0^2\rangle-\langle v^2\rangle_\mathrm{th}) G_1(t)G_1(t')\\&+\langle v^2\rangle_\mathrm{th} \left\{G_2(t)+G_2(t')-G_2(\tau)\right\}
\end{aligned}
\end{equation}
and
\begin{equation}\label{eq:xccac}   
\begin{aligned}
    \langle x_\mathrm{ac}(t)x_\mathrm{ac}(t')\rangle= \frac{\eta_\mathrm{A}^2D_\mathrm{A}}{m^2}\int_0^{t} &\left\{G_1(t'+\tau)G_3(t')\right.\\&+\left. G_1(t')G_3(t'+\tau)\right\}dt'
\end{aligned}
\end{equation}
where $G_1$, $G_2$, and $G_3$ are the Mittag-Leffler variation functions defined as:
\begin{subequations}\label{eq:MLFG123}
    \begin{align}
    \label{eq:MLFG1}
        G_1(t)&=tE_{2H,2}\left[-\left(\frac{t}{\tau_H}\right)^{2H}\right],\\
        \label{eq:MLFG2}
        G_2(t)&=\int_0^{t} G_1(t')dt'=t^2E_{2H,3}\left[-\left(\frac{t}{\tau_H}\right)^{2H}\right],\\
        \label{eq:MLFG3}
        G_3(t)&=\mathcal{L}^{-1}\left[\frac{\tau_\mathrm{A}}{(s\tau_\mathrm{A}+1)(s^2+\tau_H^{-2H} s^{2-2H})}\right](t).
    \end{align}
\end{subequations}
With the given system parameters, $\langle v^2\rangle_\mathrm{th}$ is set to be unity. Starting from these analytically derived correlator expressions, we proceed to formulate the general equations for the ensemble- and time-averaged MSDs and analyze their inherent characteristics.

\subsubsection{The ensemble-averaged MSD}\hfill

From the above expressions, the ensemble-averaged MSD is obtained as
\begin{equation}\label{eq:msdactive}
    \langle x^2(t)\rangle=\langle x_\mathrm{th}^2(t)\rangle+\langle x_\mathrm{ac}^2(t)\rangle
\end{equation} 
where the thermal part of MSD reads 
\begin{equation}\label{eq:msdFDTv0}
    \langle x_\mathrm{th}^2(t)\rangle=2\langle v^2\rangle_\mathrm{th} G_2(t)+\left(\langle v_0^2\rangle-\langle v^2\rangle_\mathrm{th} \right)G_1(t)^2
\end{equation}
and the active part is expressed as
\begin{equation}\label{eq:eamsdactive}
\langle x_\mathrm{ac}^2(t)\rangle=2\frac{\eta_\mathrm{A}^2D_\mathrm{A}}{m^2}\int_0^{t}G_1(t')G_3(t')dt'
\end{equation}
where $G_i~(i=1,2,3)$ are the aforementioned Mittag-Leffler functions shown in Eq.~\eqref{eq:MLFG123}. At the thermal (equilibrium) initial condition, the thermal MSD~\eqref{eq:msdFDTv0} increases as $\propto G_2(t)$. Thus, the thermal MSD grows with the ballistic diffusion exponent $\mu=2$ at short times ($t\ll \tau_H$) for all $H$ values and then cross-overs to the subdiffusion regime of $\sim t^{2-2H}$ in the overdamped limit ($t\gg \tau_H$). As a special case, if the system starts with $v_0=0$, the system initially performs a hyperdiffusion as $\sim t^{2+2H}$ instead of the ballistic dynamics~\cite{PRE2009Eli,PRE2021Ralf,PRE2013Ralf,PRL2010Siegle,PRE2010Siegle}. Refer to the plots of $\langle x_\mathrm{th}^2(t)\rangle$ in Fig.~\ref{figA1}(a) (Appendix~\ref{sec:appendixA}). 

The active part of MSD, $\langle x_\mathrm{ac}^2(t)\rangle$,
has three distinct scaling regimes over time. Plugging $G_3(t)\approx G_2(t)$ for $t\ll \tau_\mathrm{A}$ and $\approx\tau_\mathrm{A}G_1(t)$ for $t\gg\tau_\mathrm{A}$ into Eq.~\eqref{eq:eamsdactive}, we find that MSDs follow power-law scalings if $H\in (1/2,~3/4)$. In \textit{Case 1} where the self-propulsion time $\tau_\mathrm{A}$ is greater than the momentum relaxation time $\tau_H$---a typical condition for active colloidal systems---the MSD behaves as 
\begin{equation}\label{eq:xac2_case1}
    \langle x_\mathrm{ac}^2(t)\rangle\sim\left\{
    \begin{array}{lc}
     t^4, & ~0\leq t\ll\tau_H \\
        t^{4-4H}, & ~\tau_H\ll t\ll\tau_\mathrm{A} \\
        t^{3-4H}, & ~t\gg\tau_\mathrm{A} 
    \end{array}\right..
\end{equation}
This indicates that the active displacement exhibits a hyperdiffusion of $\sim t^4$ in the underdamped regime and, subsequently, displays a sub-ballistic superdiffusion ($\sim t^{4-4H}$) and subdiffusion ($\sim t^{3-4H}$). In \textit{Case $2$} where $\tau_H>\tau_\mathrm{A}$, the active MSD follows the scaling
\begin{equation}\label{eq:xac2_case2}
    \langle x_\mathrm{ac}^2(t)\rangle\sim\left\{
    \begin{array}{lc}
        t^4, & ~0\leq t\ll\tau_\mathrm{A} \\
        t^3, & ~\tau_\mathrm{A}\ll t\ll \tau_H\\
         t^{3-4H}, & ~t\gg\tau_H  
    \end{array}\right..
\end{equation}
In this case, the system has a complicated underdamped dynamics such that the $t^4$ hyperdiffusion is followed by a $t^3$ hyperdiffusion when $t$ is greater than $\tau_\mathrm{A}$. After these underdamped dynamics, the system suffers the subdiffusion of $\sim t^{3-4H}$ in the overdamped regime. The three distinct dynamics are separated by two important time constants of the systems, i.e., $\tau_\mathrm{A}$ and $\tau_H$. Moreover, depending on their relative magnitude, the underdamped dynamics display two distinct dynamic patterns (see the first two scalings in Eqs.~\eqref{eq:xac2_case1} \&~\eqref{eq:xac2_case2}). 
We cross-check these analytic results with the simulation. Figure~\ref{fgr3}(a) presents the plots of $\langle x_\mathrm{ac}^2(t)\rangle$ with $(\tau_H,~ \tau_\mathrm{A})=(0.357,~0.01)$ and $(0.009,~1)$. Indeed, three scaling regimes, predicted by the above theory, are clearly visible for both cases. Numerically we further find that if $\tau_\mathrm{A}\approx \tau_H$, the second regime ($t^{4-4H}$ in Case $1$ or $t^3$ in Case $2$) vanishes and the MSD grows from $\sim t^4$ to $\sim t^{3-4H}$ (not shown). Also note that if $H\geq 3/4$ the active MSD does not exhibit the power-law scaling of $\sim t^{3-4H}$ in the overdamped regime (while the underdamped power-laws remain valid), which will be explained below Eq.~\eqref{eq:xac2_long}.

We have a few comments on the underdamped dynamics in the active part of MSD. (1) When $\tau_\mathrm{A}<\tau_H$, the system has two-fold hyperballistic underdamped dynamics, $\sim t^4$ and $\sim t^3$, separated by $\tau_\mathrm{A}$ (see Eq.~\eqref{eq:xac2_case2}). We notice that these active hyperdiffusion can be understood in the context of 
the hyperdiffusion ($\sim t^{2+2H}$) shown in the equilibrium FLE starting with the initial condition of $v_0=0$.
At the timescale of $0<t\ll\tau_\mathrm{A}$, the active noise does not lose its directional correlation and acts as fGn with $H=1$ in the equilibrium FLE, so leading to the hyperdiffusion of $\sim t^4$. For $\tau_\mathrm{A}\ll t\ll\tau_H$, the active noise completely loses the directionality, which can be treated as a $\delta$-correlated noise, i.e., fGn with $H=1/2$. Thus, a hyperdiffusion of $\sim t^3$ emerges at this timescale.
(2) When $\tau_\mathrm{A}>\tau_H$, the system has a single underdamped dynamics of $\sim t^4$ for $t\ll \tau_H$ by the same reasoning explained in (1). A very interesting overdamped regime occurs for this case at the timescale of $\tau_H\ll t\ll \tau_\mathrm{A}$. Here the system's momentum relaxation is sufficiently reached while the active noise is persistently imposed on the system like a constant force. The resultant active dynamics turns out to be the sub-ballistic superdiffusion with the MSD of $\sim t^{4-4H}$, capturing the viscoelastic property of the environment. If the medium is viscous ($H=1/2$), the system shows ballistic dynamics due to the very persistent active force. If the medium is elastic, the active force is fully stored into the elastic energy without any kinetic energy ($\sim t^0$). In a viscoelastic medium between these two limits, the system displays superdiffusive ($1/2<H<3/4$), Fickian ($H=3/4$), and subdiffusive ($3/4<H<1$) dynamics depending on the Hurst exponent. 
 
In the long-time regime where $t\gg\tau_H$ and $\tau_\mathrm{A}$, Eq.~\eqref{eq:eamsdactive} yields
\begin{equation}
\label{eq:msdlong}
\langle x_\mathrm{ac}^2(t)\rangle\approx\frac{2k_B\mathcal{T}D_\mathrm{A}\tau_\mathrm{A}\sin{(\pi (2-2H))}}{\gamma_0\pi\Gamma(2H-1)\Gamma(3-2H)}\int_0^t y^{2-4H}dy.
\end{equation} 
This shows that the long-time active dynamics are classified into the following three cases: 
\begin{equation}\label{eq:xac2_long}
    \langle x_\mathrm{ac}^2(t)\rangle\approx A\times\left\{
    \begin{array}{lc}
        \frac{1}{3-4H}t^{3-4H}, & \frac{1}{2}<H<\frac{3}{4} \\ 
        \ln{t}, & H=\frac{3}{4}  \\ 
        \mathrm{const}., & \frac{3}{4}<H<1  
    \end{array}\right.
\end{equation}
with $A=\frac{2k_B\mathcal{T}D_\mathrm{A}\tau_\mathrm{A}\sin{(\pi (2-2H))}}{\gamma_0\pi\Gamma(2H-1)\Gamma(3-2H)}$.
We note that Eq.~\eqref{eq:xac2_long} is consistent with the results reported in some previous studies on FDT-violating overdamped systems~\cite{PRE1996Porra,WANG1999,PRE2006Kwok}. For example, an overdamped nonequilibrium FLE having a power-law memory kernel ($\sim t^{-\alpha}$) conjugated with a fGn of autocovariance decaying as $\sim t^{-\beta}$ exhibits $\langle x^2(t)\rangle\sim t^{2\alpha-\beta}$ for $\alpha>\beta/2$, $\ln t$ at $\alpha=\beta/2$, and a $\mathrm{constant}$ for $\alpha<\beta/2$ ~\cite{PRE1996Porra,WANG1999,PRE2006Kwok}. In Refs.~\cite{PRE1996Porra,SoftMatter2020Joo,Han2023,Michael2000PRL,PRE2005BaoOUN,Softmatter2017Vandebroek,Grimm2018Softmatter,Sakaue2017Softmatter}, a nonequilibrium FLE with an FDT-violating noise having a finite correlation time (e.g., an exponential function) was investigated. The study demonstrated that $\langle x^2(t)\rangle\sim t^{2\alpha-1}$ for $1/2<\alpha<1$, $\ln t$ at $\alpha=1/2$, and $\mathrm{a~constant}$ for $0<\alpha<1/2$.

In Fig.~\ref{fgr3}(b), we plot the total MSD, $\langle x^2(t)\rangle=\langle x_\mathrm{th}^2(t)\rangle+\langle x_\mathrm{ac}^2(t)\rangle$, for varying $D_\mathrm{A}$ and $v_0$. In the absence of the active noise ($D_\mathrm{A}=0$), the system's dynamics follows that of the equilibrium FLE. Here, the total MSD corresponds to $\langle x_\mathrm{th}^2(t)\rangle$, where the short-time dynamics are either ballistic if $v_0\neq0$ or hyperdiffusive ($\sim t^{2+2H}$) if $v_0=0$. At the overdamped timescale, they crossover to subdiffusive dynamics ($\sim t^{2-2H}$). 
When a strong noise is introduced ($D_\mathrm{A}=400$), the scaling properties of $\langle x^2(t)\rangle$ mostly follow those of $\langle x_\mathrm{ac}^2(t)\rangle$. However, the short-time regime is altered. If $v_0\neq0$, the ballistic scaling ($\sim t^2$) from the thermal MSD dominates over the hyperdiffusion of the active component ($\sim t^4$). Accordingly, the full MSD initially increases with $\sim t^2$, then has a transient accelerating dynamics ($\sim t^4$), and enters the scaling regimes of the superdiffusion $\sim t^{3/2}$ and the active subdiffusion $\sim t^{1/2}$ as time progresses. This corresponds to the scenario described by Eq.~\eqref{eq:xac2_case1} with $\tau_\mathrm{A}>\tau_H$. When $v_0=0$, the MSD grows with $\sim t^{2+2H}$, accompanied with a short cross-over ($\sim t^4$), and exhibits the regimes of $\sim t^{3/2}$ and $\sim t^{1/2}$ with increasing time. We note that the active subdiffusion persists until a cross-over time 
\begin{equation}\label{eq:taustar}
    \tau^*\approx O(D_\mathrm{A}^{1/(2H-1)}).
\end{equation}
For example, when $H=3/4$ and $D_\mathrm{A}\sim O(10^2)$, $\tau^*$ is estimated to be $O(10^6)$ with other constant factors. For $t\gg \tau^*$, the thermal motion eventually dominates in EA MSD, and the MSD exhibits the thermal subdiffusion ($\sim t^{2-2H}$).

\begin{figure*}
    \centering
    \includegraphics[width=1\textwidth]{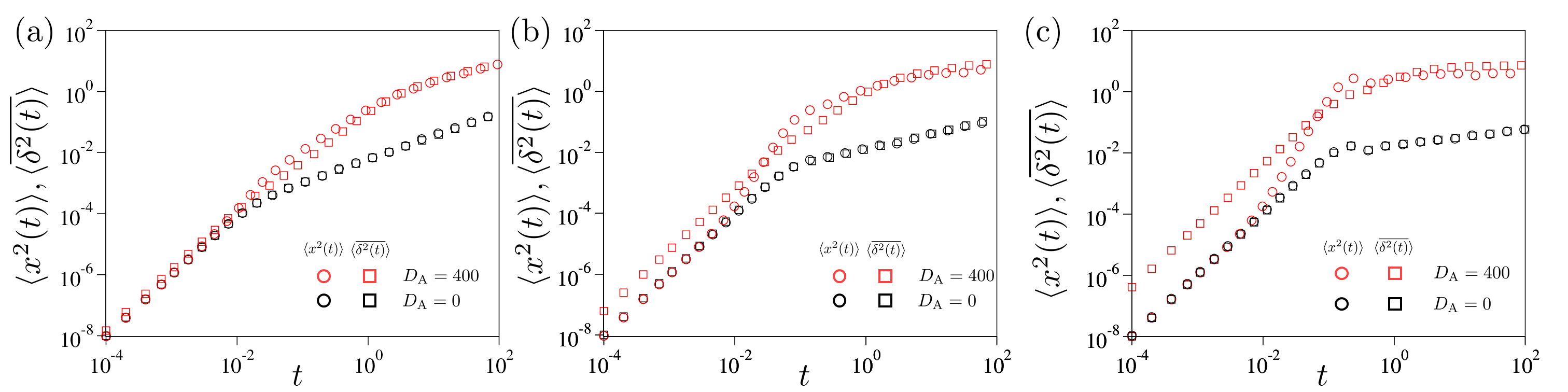}
\caption{ The comparison of $\langle x^2(t)\rangle$ and $\langle \overline{\delta^2(t)}\rangle$ for three $H$ values for Case 1 ($\tau_\mathrm{A}>\tau_H$). In each plot, the simulation results for $D_\mathrm{A}=0$ and $400$ are present. (a) $H=5/8$ with $\tau_\mathrm{A}=1$ and $\tau_H=0.009$. (b) $H=3/4$ with $\tau_\mathrm{A}=1$ and $\tau_H=0.03$. (c) $H=7/8$ with $\tau_\mathrm{A}=1$ and $\tau_H=0.064$. 
For all panels, we use an initial condition of $\langle v_0^2\rangle=1$. }
\label{fgr4}
\end{figure*}
\subsubsection{The time-averaged MSD} \hfill 

Instead of the ensemble-averaged MSD above, we can alternatively define an MSD from a single trajectory $x(t)$ in the time-averaging sense:
\begin{equation}
\overline{\delta^2(t,T)}=\frac{1}{T-t}\int_0^{T-t} [x(t'+t)-x(t')]^2 dt'.
\end{equation} 
In this expression, $t$ acts as time lag and $T$ is the total measurement time. Given the fact that the active FLE~\eqref{eq:FLEwoFDT2} is a stationary process, here we define and study the time-averaged (TA) MSD $\overline{\delta^2(t)}=\lim_{T \to\infty}\overline{\delta(t,T)}$ for sufficiently long trajectories. Further, we define the ensemble-average of $\overline{\delta^2(t)}$ over many trajectories, which is expressed as
\begin{equation}
\label{eq:tamsd2}
\langle \overline{\delta^2(t)}\rangle =\lim_{T\to\infty}\frac{2}{T-t}\int_0^{T-t} [\langle x^2(t')\rangle -\langle x(t'+t)x(t')\rangle] dt'.
\end{equation} 
In this integral, we utilize the stationary parts of $\langle x^2(t)\rangle$ and $\langle x(t'+t)x(t')\rangle$ because the effect from the $v_0$-dependent terms in Eq.~\eqref{eq:tamsd2} becomes negligible if $T$ is sufficiently long. 

The TA MSD is also expressed as the superposition of the equilibrium and active components, i.e, $\langle \overline{\delta^2(t)}\rangle=\langle \overline{\delta_\mathrm{th}^2(t)}\rangle+\langle \overline{\delta_\mathrm{ac}^2(t)}\rangle$. The thermal part of TA MSD is given by~\cite{PRE2009Eli,PRE2010Jeon}
\begin{equation}\label{eq:tamsdth}
    \langle \overline{\delta_\mathrm{th}^2(t)}\rangle=2\langle v^2\rangle_\mathrm{th} G_2(t),
\end{equation} 
which precisely corresponds to the thermal part of EA MSD~\eqref{eq:msdFDT} under the equilibrium condition. The active part of TA MSD is approximated as:
\begin{equation}\label{eq:tamsdactive}
\begin{aligned}
   \frac{\langle \overline{\delta_\mathrm{ac}^2(t)}\rangle }{2A'}\approx&\int_t^{\infty} G_1(t'-t)G_3(t'-t)dt'+\int_0^{\infty} G_1(t')G_3(t')dt'\\&-\int_0^{\infty} \left[G_1(t'+t)G_3(t')+ G_1(t')G_3(t'+t)\right]dt' 
\end{aligned}
\end{equation} 
where $A'=\eta_\mathrm{A}^2D_\mathrm{A}/m^2$, and $G_1(t)$ and $G_3(t)$ are the Mittag-Leffler variation functions, Eqs.~\eqref{eq:MLFG1} \&~\eqref{eq:MLFG3}, respectively. By utilizing the Taylor expansion of Eq.~\eqref{eq:tamsdactive}, we deduce the scaling properties of the TA MSD at $t\to0$ and $t\to\infty$. In the former case, $\langle \overline{\delta_\mathrm{ac}^2(t)}\rangle$ behaves as
\begin{equation}\label{eq:tamsdac0}
\begin{aligned}
    \langle \overline{\delta_\mathrm{ac}^2(t)}\rangle&\approx 2\frac{\eta_\mathrm{A}^2D_\mathrm{A}}{m^2}\left(\int_0^\infty F_1(t')F_2(t')dt'\right)t^2\\&=\left(\langle\overline{v^2}\rangle_\mathrm{st}-\langle v^2\rangle_\mathrm{th}\right) t^2,
\end{aligned}
\end{equation} 
indicating that the active part of TA MSD always exhibits a ballistic scaling, regardless of the values of $H$. In contrast, the long-time overdamped asymptotic for $ t\gg\tau_\mathrm{A},\tau_H$ is highly dependent on the values of $H$. 
(1) For the range $1/2<H<3/4$, we obtain
\begin{equation}\label{eq:tamsdlong1}
\langle \overline{\delta_\mathrm{ac}^2(t)}\rangle
\approx-2A\frac{\Gamma(4H-3)\Gamma(2-2H)}{\Gamma(2H-1)}
t^{3-4H}.
\end{equation}
Here $A=\frac{2k_B\mathcal{T}D_\mathrm{A}\tau_\mathrm{A}\sin{(\pi (2-2H))}}{\gamma_0\pi\Gamma(2H-1)\Gamma(3-2H)}$ is the same prefactor that also appears in Eq.~\eqref{eq:xac2_long} and $\Gamma(4H-3)<0$.
It is noted that the active part of TA MSD exhibits the same subdiffusive power-law exponent $3-4H$ as the corresponding EA MSDs~\eqref{eq:xac2_case1} \&~\eqref{eq:xac2_case2}. However, the amplitudes of the EA and TA MSDs are not identical. 
(2) In the case of $H=3/4$, the TA MSD is obtained to be
\begin{equation}\label{eq:tamsdlong2}
\frac{\langle \overline{\delta_\mathrm{ac}^2(t)}\rangle}{A}\approx 2\gamma_\mathrm{E}+2\ln{t}+\frac{\Gamma'(2-2H)}{\Gamma(2-2H)}+\frac{\Gamma'(2H-1)}{\Gamma(2H-1)}
\end{equation}
where $\gamma_\mathrm{E}=0.5772$ is the Euler's constant and $\Gamma'(z)=d\Gamma(z)/dz$. It is important to note that the active TA MSD differs from $\langle x_\mathrm{ac}^2(t) \rangle$ by a factor of $2$ in front of the $\ln t$ term and also by other extra constants. (3) For the range $3/4<H<1$, the active TA MSD follows that of a confined diffusion, where the long-time asymptotic value of the TA MSD is twice the value of the counterpart EA MSD. The disagreement between the EA and TA MSDs is further analyzed in detail in Sec.~\ref{sec:EB}.

Figure~\ref{fgr3}(c) illustrates the simulated $\langle \overline{\delta_\mathrm{ac}^2(t)}\rangle$ for $H=5/8$, considering three cases: $\tau_H>\tau_\mathrm{A}$, $\tau_H\approx\tau_\mathrm{A}$, and $\tau_H<\tau_\mathrm{A}$. In all three cases, the active part of TA MSD exhibits the expected two distinct power-law regimes ($\sim t^2$ and $\sim t^{1/2}$) over time. This is in contrast with the complicated scaling behaviors observed in $\langle x_\mathrm{ac}^2(t)\rangle$. Notably, the initial hyperdiffusion ($\sim t^4$) observed in the EA counterpart is replaced by ballistic diffusion, and the accompanied transient scaling regimes ($\sim t^3$ or $\sim t^{4-4H}$) vanish. 
Furthermore, the simulation confirms that the long-time dynamics occur for $t\gtrsim \mathrm{max}(\tau_H,~\tau_\mathrm{A})$. 

In Fig.~\ref{fgr3}(d), we present the total TA MSD $\langle \overline{\delta^2(t)}\rangle$ for increasing $D_\mathrm{A}$ while keeping $\tau_\mathrm{A}$ and $\tau_H$ fixed.
At $D_\mathrm{A}=0$, the TA MSD exhibits only two scaling regimes: an initial ballistic dynamics for $t\lesssim \mathrm{max}(\tau_H,\tau_\mathrm{A})$, followed by the thermal subdiffusion characterized by a power-law of $\sim t^{2-2H}$ beyond the cross-over time. However, in the presence of the active noise ($D_\mathrm{A}\neq 0$), the scaling behaviors become more complex. After the ballistic regime, a transient regime of superdiffusion emerges in the time window of $\tau_H<t<\tau_\mathrm{A}$. In this plot, we observe an apparent anomalous exponent of approximately $1.7$, which arises from the superposition of $\langle \overline{\delta_\mathrm{ac}^2(t)}\rangle\sim t^2$ and $\langle \overline{\delta_\mathrm{th}^2(t)}\rangle\sim t^{2-2H}$. After this regime, the TA MSD exhibits active subdiffusion with a scaling relation of $\sim t^{3-4H}$ (e.g., see $t^{1/2}$ at $D_\mathrm{A}=400$). For $t\gg \tau^*$, the TA MSD is expected to enter the final regime of the thermal subdiffusion ($\sim t^{2-2H}$).

Figure~\ref{fgr4} compares the EA and TA MSDs for different values of the Hurst exponent: (a) $H=5/8$, (b) $3/4$, and (c) $7/8$ for Case $1$ ($\tau_\mathrm{A}>\tau_H$). In each panel, the EA and TA MSDs are simulated with $D_\mathrm{A}=0$ and $400$ with an initial condition of $\langle v_0^2\rangle=1$. As expected, the active-free system ($D_\mathrm{A}=0$) has identical EA and TA MSDs, demonstrating the ergodicity of the equilibrium FLE process. On the contrary, the EA and TA MSDs deviate from each other for the active FLEs ($D_\mathrm{A}=400$). These discrepancies signify that the active FLE~\eqref{eq:FLEwoFDT2} is non-ergodic within our observation time window; the discrepancies become more pronounced as $H$ increases. In Fig.~\ref{figureC1}, we present the EA and TA MSDs for Case $2$ ($\tau_H>\tau_\mathrm{A}$), with the same values of $H$. Analogously to Fig.~\ref{fgr4}, the two MSDs exhibit increasing discrepancies with larger values of $H$. However, a difference is that the TA MSDs are always greater than the EA MSDs across all times. This observation will be further examined below in Sec.~\ref{sec:EB}. 

We emphasize that the observed non-ergodic behavior arises from the combined influence of the viscoelastic memory and underdamped dynamics in the system. The conventional AOUP model, which neglects the memory term in the active FLE~\eqref{eq:FLEwoFDT2}, is ergodic regardless of the active motion. In the active FLE model where the viscoelastic effect is incorporated, the active component reveals ultraweak ergodicity breaking in the sense that the EA and TA MSDs exhibit the same scaling form but differ only in their amplitudes, as Eqs.~\eqref{eq:xac2_long},~\eqref{eq:tamsdlong1}, \&~\eqref{eq:tamsdlong2} show above.
\begin{figure*}

\centering\includegraphics[width=0.75\textwidth]{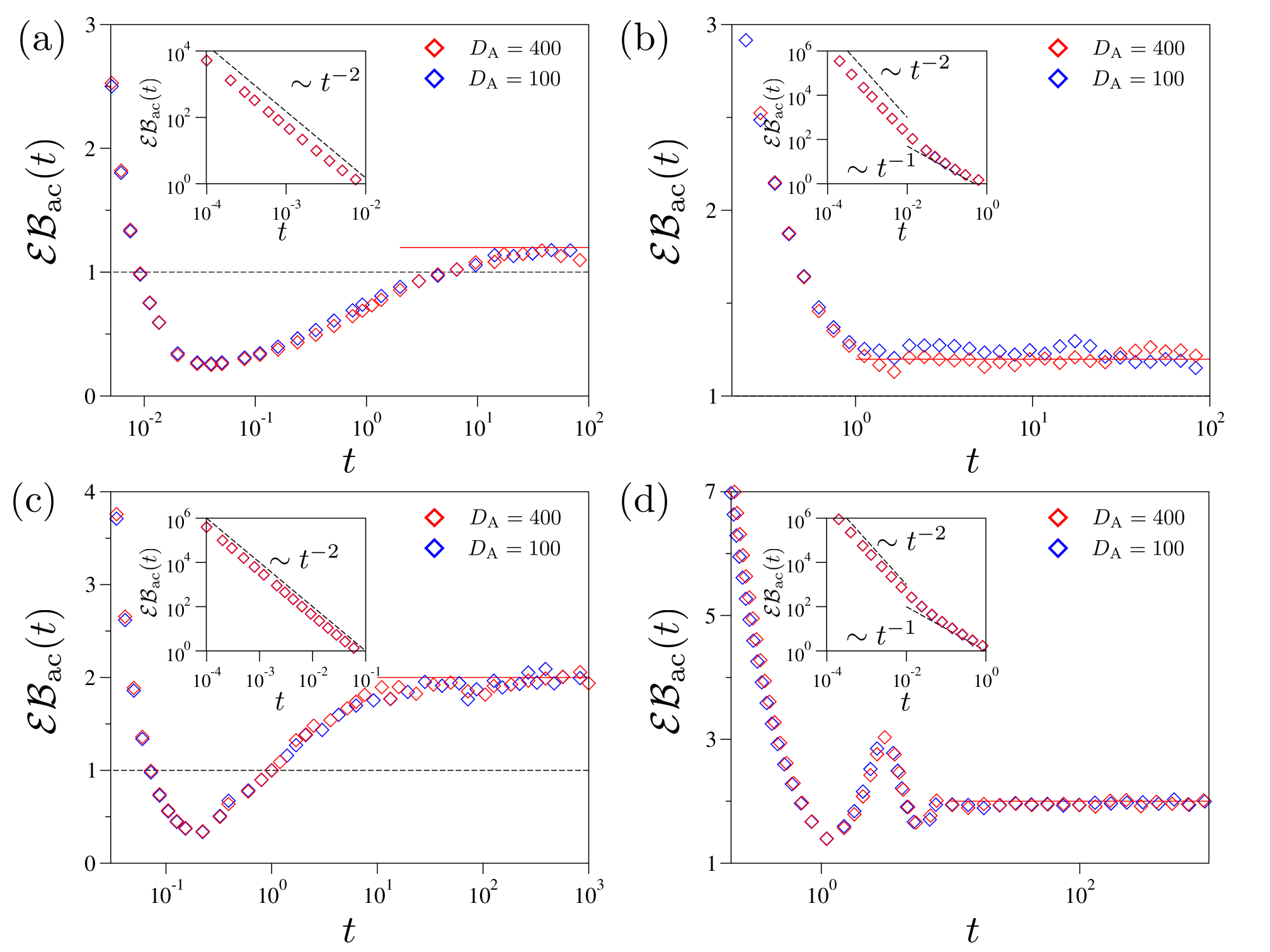}  
    
\caption{The time evolution of $\mathcal{EB}_\mathrm{ac}(t)$ for various conditions. Left panels are the results for Case $1$ ($\tau_\mathrm{A}>\tau_H$), and right panels for Case $2$ ($\tau_H>\tau_\mathrm{A}$). (a) $\mathcal{EB}_\mathrm{ac}(t)$ with $\tau_\mathrm{A}=1$ and $\tau_H=0.009$ ($H=5/8$). Simulation results (symbol) for $D_\mathrm{A}=100$ and $400$ approaches to $\mathcal{EB}_{\mathrm{ac},\infty}$ (the red solid line). $\mathcal{EB}_{\mathrm{ac},\infty}=1.2$ for $H=5/8$. Inset: $\mathcal{EB}_\mathrm{ac}(t)$ in the short-time regime with a power-law scaling of $\sim t^{-2}$. 
(b) $\mathcal{EB}_\mathrm{ac}(t)$ with $\tau_\mathrm{A}=0.01$, and $\tau_H=0.36$ ($H=5/8$). Inset: $\mathcal{EB}_\mathrm{ac}(t)$ in the short-time regime with a power-law scaling of $\sim t^{-2}$ and $\sim t^{-1}$ 
(c) $\mathcal{EB}_\mathrm{ac}(t)$ with $\tau_\mathrm{A}=1$, and $\tau_H=0.064$ ($H=7/8$). $\mathcal{EB}_{\mathrm{ac},\infty}=2$ is depicted for $H=7/8$. Inset: $\mathcal{EB}_\mathrm{ac}(t)$ in the short-time regime. 
(d) $\mathcal{EB}_\mathrm{ac}(t)$ with $\tau_\mathrm{A}=0.01$, and $\tau_H=0.88$ ($H=7/8$). Inset: $\mathcal{EB}_\mathrm{ac}(t)$ in the short-time regime. } 
\label{fgr5}
\end{figure*}
\begin{figure*}
\centering\includegraphics[width=0.75\textwidth]{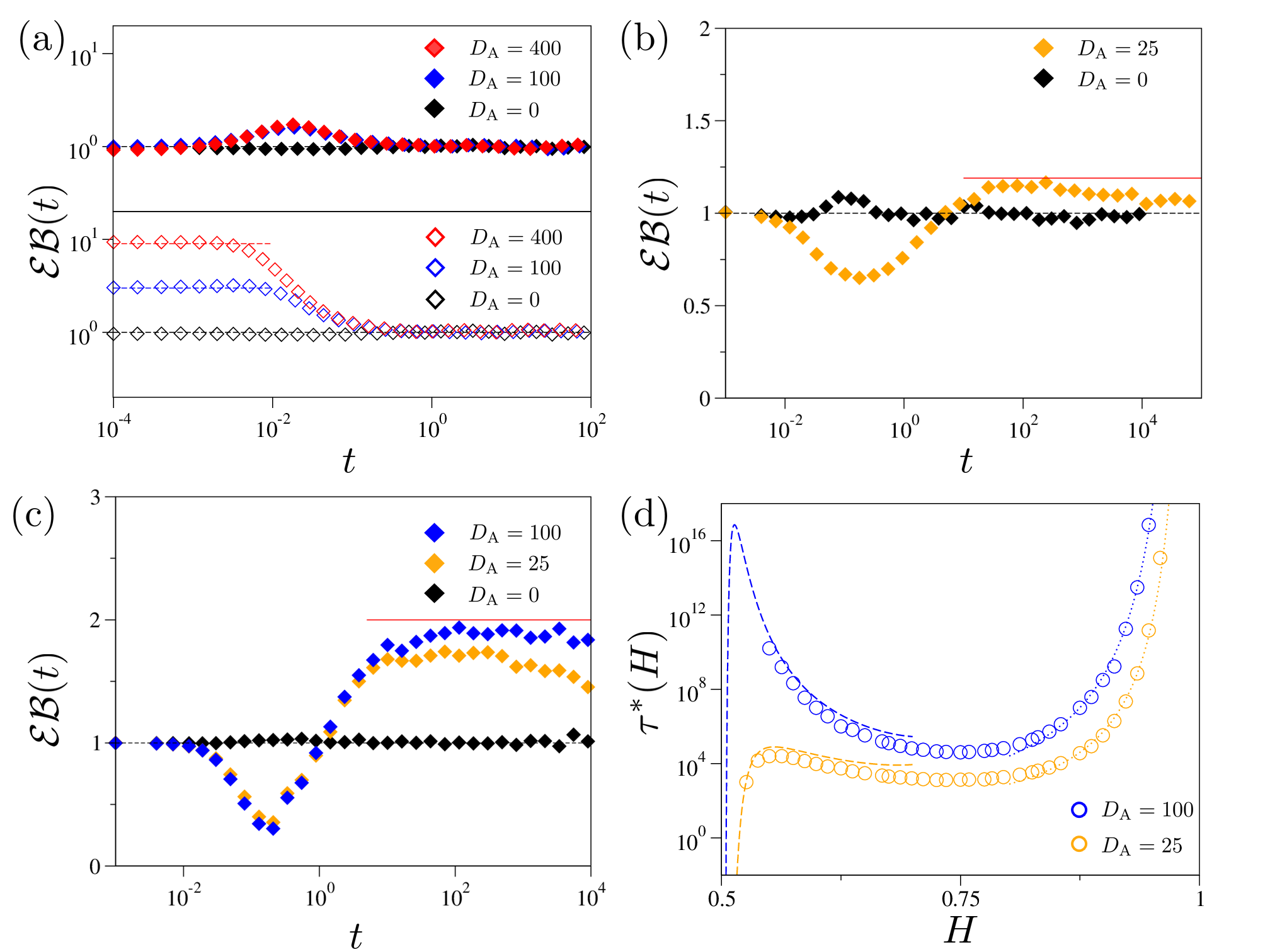} 
    
\caption{The profile of $\mathcal{EB}(t)$ and $\tau^*(H)$. 
(a) $\mathcal{EB}(t)$ for the active LE model (i.e., the conventional AOUP model~\cite{PRL2016AOUP,PRE2019Bonilla,AOUP2022Lowen}) with $\tau_\mathrm{A}=1$ \& $\tau_{1/2}=0.01$. Two distinct cases are simulated with different initial conditions. Upper panel: the stationary initial condition $\langle v_0^2 \rangle=\langle \overline {v^2}\rangle_\mathrm{st}$. 
Lower panel: a thermal initial condition
$\langle v_0^2 \rangle=1$. 
(b) $\mathcal{EB}(t)$ for an active FLE system with $H=5/8$ and $\tau_\mathrm{A}\gg\tau_H$ (Case $1$). The system parameters are: $\tau_\mathrm{A}=1$, $\tau_H=0.009$, $D_\mathrm{A}=0$ and $25$. 
(c) $\mathcal{EB}(t)$ with $H=7/8$ for Case $1$. The simulation parameters are: $\tau_\mathrm{A}=1$, $\tau_H=0.064$, $D_\mathrm{A}=0$, $25$, and $100$. 
(d) The numerically estimated $\tau^*$ (circle) as a function of $H$. 
The dashed lines: Eq.~\eqref{tau*1} in the regime of $1/2<H<3/4$. The dotted lines: Eq.~\eqref{tau*2} in the regime of for $3/4<H<1$, where the saturated value of the active part MSD is empirically evaluated to $C=\langle x_\mathrm{ac}^2\rangle(t=10^2)$ at $H=7/8$.
} 
\label{fgr6}
\end{figure*}

\subsection{Ergodicity-breaking parameter}\label{sec:EB}

The analysis of MSDs in the active FLE model reveals that the system exhibits a non-ergodic behavior at certain timescales, and the propensity of non-ergodicity strongly depends on the elapsed time $t$ and Hurst exponent $H$. To quantify the extent of ergodicity breaking, we introduce the ergodicity-breaking (EB) parameter, given by the expression~\cite{PRL2013EBRalf}:
\begin{equation}\label{eq:EB}
    \mathcal{EB}(t)=\frac{\langle\overline{\delta^2(t)}\rangle}{\langle x^2(t)\rangle}.
\end{equation} 

Using the EB parameter, we begin by investigating the non-ergodic behavior of the active component dynamics in the active FLE system. In Fig.~\ref{fgr5}, we present $\mathcal{EB}_\mathrm{ac}(t)=\langle\overline{\delta_\mathrm{ac}^2(t)}\rangle/\langle x_\mathrm{ac}^2(t)\rangle$ for $H=5/8$ (a, b) and $H=7/8$ (c, d). For each $H$ value, we compare the case of $\tau_\mathrm{A}>\tau_H$ (Left panels) with the other one of $\tau_H>\tau_\mathrm{A}$ (Right panels). The EB parameters evidently show that the active viscoelastic dynamics are non-ergodic, and the degree of ergodicity breaking significantly changes with time. In Case $1$ (Left panels), the system exhibits a strong disparity between the two MSDs in the underdamped regime where $\mathcal{EB}_\mathrm{ac}(t)$ decays as $\sim t^{-2}$ for $t\lesssim \tau_H$ (see the inset plots therein). The EB parameter decreases below unity in the intermediate timescale of $\tau_H\lesssim t\lesssim \tau_\mathrm{A}$. In the final regime of $t\gg \tau_\mathrm{A}$, $\mathcal{EB}_\mathrm{ac}(t)$ reaches a stationary value, $\mathcal{EB}_\mathrm{ac,\infty}(\neq 1)$ and the active component displays \textit{ultraweak ergodicity-breaking} (UWEB), as aforementioned. We analytically obtain $\mathcal{EB}_\mathrm{ac,\infty}(\neq 1)$ from the formal expressions for the two MSDs, which reads
\begin{equation}\label{eq:EB_ac_inf} \mathcal{EB}_{\mathrm{ac},\infty}\approx\left\{\begin{array}{lc}
     1 & H=\frac{1}{2}  \\
    2\frac{\Gamma(4H-2)\Gamma(2-2H)}{\Gamma(2H-1)} & \frac{1}{2}<H<\frac{3}{4} \\
    2 & H=\frac{3}{4} \\
    2 & \frac{3}{4}<H<1 
\end{array}\right..
\end{equation} 
Here, the case of $H=1/2$ refers to the active LE, i.e., the AOUP in a viscous medium. For the active viscoelastic system with $H\in (1/2,~1)$, the EB parameter satisfies $\mathcal{EB}_{\mathrm{ac},\infty}\in (1,~2]$ and increases monotonically with $H$ (see Fig.~\ref{figureC2}). Figure~\ref{fgr5} indeed shows that the theoretical prediction [Eq.~\eqref{eq:EB_ac_inf}, the solid line] explains successfully the simulated $\mathcal{EB}_{\mathrm{ac},\infty}$. Surprisingly, the results suggest that the extent of ergodicity breaking only depends on the viscoelastic memory parameter $H$, but the noise strength does not contribute to the ergodicity breaking. In Case $2$ (Right panels), the system exhibits time-dependent ergodicity breaking as well, however, of which pattern is clearly distinguished from Case $1$. In this case, the ensemble-averaged MSD has two distinct underdamped scaling laws [see Eq.~\eqref{eq:xac2_case2}], which makes the two power-law scaling of $\sim t^{-2}$ and $\sim t^{-1}$ in $\mathcal{EB}_{\mathrm{ac}}(t)$ for $t<\tau_H$; see the insets. In the overdamped regime, the EB parameter is eventually saturated to $\mathcal{EB}_{\mathrm{ac},\infty}$ predicted by Eq.~\eqref{eq:EB_ac_inf}. Intriguingly, $\mathcal{EB}_{\mathrm{ac}}$ approaches the stationary value with oscillation when $H$ is close to $1$. 

In Fig.~\ref{fgr6}, we examine the EB parameter for the total dynamics of the active FLE~\eqref{eq:FLEwoFDT2}, i.e., the sum of the thermal and active components. 
Before investigating the active FLE system, we examine the active LE model, the so-called conventional AOUP model that neglects the viscoelastic memory~\cite{PRL2000Libchaber,JCP2020Lowen,IOP2021Nguyen,PRE2021Martin,PRE2019Bonilla,PCCP2021abp}, corresponding to the active FLE~\eqref{eq:FLEwoFDT2} with $H\to 1/2$. In Fig.~\ref{fgr6}(a, upper), we evaluate $\mathcal{EB}(t)$ for this model with initial velocities taken from the stationary velocity distribution ($\langle v_0^2\rangle=\langle\overline{v^2}\rangle_\mathrm{st}$). The simulation results show that the EB parameter, starting from unity at short times, deviates from it in the timescale of $\tau_{1/2}\lesssim t\lesssim\tau_\mathrm{A}$ (here $\tau_{1/2}=m/\gamma_0$), and eventually recovers unity. This pattern remains the same regardless of the noise strength. These results demonstrate that the memory-free AOUP system is ergodic~\cite{PRE2018Cherstvy,JPA2020Wang}, even though it violates FDT. A noteworthy feature is that the system suffers \textit{transient} ergodicity breaking on the intermediate timescale, even under the stationary initial condition. This behavior is clearly distinguished from the ordinary underdamped LE (the case at $D_\mathrm{A}=0$ in Fig.~\ref{fgr6}(a)), where $\mathcal{EB}(t)=1$ for all times. This transient ergodicity breaking is attributed to the fact that the AOUP model is subject to two independent noise sources of which stationary properties are distinguished. This indicates that either the thermal or active components or both of them will start initially in a non-stationary state. The thermal and active parts are then separately relaxed to their stationary states with the timescales of $\tau_H$ (thermal) and $\tau_\mathrm{A}$ (active). In the lower panel of Fig.~\ref{fgr6}(a), we plot $\mathcal{EB}(t)$ with a non-stationary initial condition $\langle v_0^2\rangle=1$ for the same AOUP model.
As will be explained further below, in this case, the EB parameter starts from the short-time asymptotic value $\mathcal{EB}_0\approx \langle\overline{v^2}\rangle_\mathrm{st}/\langle v_0^2\rangle$. Because $\langle\overline{v^2}\rangle_\mathrm{st}$ increases proportionally to $D_\mathrm{A}$, the short-time EB, $\mathcal{EB}_0$, can be arbitrarily large. 
As the AOUP system is ergodic, the EB parameters then monotonically decay to $\mathcal{EB}=1$ after $t\gtrsim \tau_\mathrm{A}$.

Now we investigate $\mathcal{EB}(t)$ for the active viscoelastic system where the system has a power-law decaying memory kernel. Figures~\ref{fgr6}(b) and~\ref{fgr6}(c) shows the variation of the EB parameters for $H=5/8$ (b) and $7/8$ (c) with the stationary initial condition and $\tau_\mathrm{A}>\tau_H$.
For the active FLE system, importantly, $\mathcal{EB}(t)$ experiences four distinct regimes over time. The first regime is the initial-condition dominating regime in the underdamped domain ($t\ll\tau_H$, $\tau_\mathrm{A}$), where $\mathcal{EB}(t)$ has the following asymptotic behavior:
\begin{equation}
\label{eq:Bshortactive}
    \mathcal{EB}_0=\left\{\begin{array}{ll}
\langle \overline{v^2}\rangle_\mathrm{st}/\langle v_0^2\rangle,& v_0\neq 0\\
    t^{-2H}, & v_0= 0
\end{array}\right..
\end{equation}
In Figs.~\ref{fgr6}(b) and~\ref{fgr6}(c), $\mathcal{EB}_0$ is unity under the stationary initial condition. Eq.~\eqref{eq:Bshortactive} indicates that at a non-stationary condition $\mathcal{EB}_0$ can be arbitrary and its value strongly depends on the initial velocity, $D_\mathrm{A}$, and $\tau_\mathrm{A}$. As a special case, when $v_0=0$, $\mathcal{EB}_0$ has no asymptotic value but decays algebraically with an exponent of $-2H$. 

After this regime, $\mathcal{EB}(t)$ 
enters the transient ergodicity-breaking regime in the time window of $\tau_H\lesssim t\lesssim \tau_\mathrm{A}$, where the EB parameter significantly varies over time. Beyond this regime, the system approaches the \textit{active dynamics-dominating ergodicity-breaking} regime. Here, $\mathcal{EB}(t)$ is saturated to a time-independent constant that deviates from unity. If the active noise is sufficiently stronger than the thermal, the constant is approximated to $\mathcal{EB}_{\mathrm{ac},\infty}$ explained in Eq.~\eqref{eq:EB_ac_inf}. Then, the property of $\mathcal{EB}(t)$ in this regime follows that of the active component of the active FLE studied above. 
This ergodicity-breaking regime persists up to $t\approx\tau^*$, at which the thermal MSD becomes comparable with the active counterpart [see Eq.~\eqref{eq:taustar}]. For $t> \tau^*$, the thermal dynamics dominates over the active part and, thus, $\mathcal{EB}(t)$ will converge to unity as $t\to\infty$. Therefore, the active FLE~\eqref{eq:FLEwoFDT2} is, in principle, ergodic in the infinite-time limit. 

It is noteworthy that despite the ergodic nature of the FLE model, the presence of viscoelasticity gives rise to a remarkably long active dynamics-dominating ergodicity-breaking state in the time window of $\tau_\mathrm{A}\lesssim t\lesssim \tau^*$. To quantify this behavior,
we plot $\tau^*(H)$ in Fig.~\ref{fgr6}(d) by numerically solving $\langle x_\mathrm{th}^2(\tau^*)\rangle=\langle x_\mathrm{ac}^2(\tau^*)\rangle$. 
The numerical data (circles) are explained by two analytic approximations (depicted as dashed \& dotted lines). In the regime of $1/2<H<3/4$, the approximate expression for $\tau^*(H)$ can be obtained by equating Eqs.~\eqref{eq:msdlimit} and~\eqref{eq:xac2_long}, yielding: 
\begin{equation}\label{tau*1}
\tau^*(H)=\left(\frac{D_\mathrm{A}\tau_\mathrm{A}\sin(\pi(2-2H))}{\pi(3-4H)}\right)^{1/(2H-1)}~\hbox{for}~H\in (1/2,~3/4).
\end{equation}
This expression is depicted by the dotted lines. In the regime of $3/4<H<1$, the active MSD follows that of a confined diffusion [see Eq.~\eqref{eq:xac2_long}]. By approximating the active MSD as a constant $C$ independent of $H$ (see further information in the Caption), we obtain the following approximation:
\begin{equation}\label{tau*2}
\tau^*(H)=\left(\frac{C\Gamma(2H-1)\Gamma(3-2H)}{2\gamma_0}\right)^{1/(2-2H)}~\hbox{for}~H\in (3/4,~1).
\end{equation}
The theoretical curve~\eqref{tau*2} is depicted by the dashed lines in the plot. Both analytic curves are shown to explain the numerical data successfully. The plot reveals a non-monotonic dependence of $\tau^*$ on $H$. For $1/2<H<3/4$, $\tau^*$ exhibits a local maximum around $H\sim 1/2$, and its peak value increases significantly with increasing $H$. Beyond this point, $\tau^*$ monotonically decreases and reaches a local minimum point at $H=3/4$. In the domain of $3/4<H<1$, $\tau^*$ rapidly increases with $H$ and diverges in the limit as $H\to1$. The same trend is observed at a higher $D_\mathrm{A}$ while the magnitude of $\tau^*$ is sensitive to $D_\mathrm{A}$. 

The behaviors of $\tau^*(H)$ suggest that observing the final ergodic regime where $\mathcal{EB}\to 1$ is nontrivial for the active FLE system. This effect becomes particularly prominent for systems with $H>3/4$, where $\tau^*$ diverges as $H\to 1$. As an illustration, Fig.~\ref{fgr6}(c) displays the plot of $\mathcal{EB}(t)$ for $H=7/8$ up to $t=10^4$. Once the system enters the active dynamics-dominating ergodicity-breaking state, characterized by $\mathcal{EB}_{\mathrm{ac},\infty}\approx 2$, the EB parameter does not decay within the observed time window. By estimating $\tau^*$ to be approximately $10^8$ for $D_\mathrm{A}=100$, we can infer that this ergodicity-breaking state will persist at least up to $t\sim 10^8$. Therefore, it is worth noting that while the active FLE system with $1/2<H<1$ can exhibit ergodic behavior in the infinite measurement time, the presence of viscoelastic long-time memory introduces effectively ergodicity-breaking phenomena within the finite measurement window. This indicates that the interplay between active noise and viscoelasticity plays a crucial role in shaping the dynamics and the observed ergodic properties of the system.




\section{Concluding remarks}\label{sec4}

In this study, we have performed a comprehensive investigation into the under- and overdamped dynamics and ergodicity-breaking phenomena for the active FLE systems modeled by Eq.~\eqref{eq:FLEwoFDT2}. A representative example within this class of models is the nonequilibrium diffusion of active Ornstein-Uhlenbeck particles (AOUPs) embedded in a viscoelastic medium. Our approach involved analytic calculations of various dynamic quantities, including the velocity and position autocorrelation functions, ensemble- and time-averaged MSDs, as well as the ergodicity-breaking parameters. 

We found that the active FLE systems are decomposed into two independent dynamic components, i.e., the thermal and active parts. On the one hand, the former follows an equilibrium FLE~\cite{PRE2001Lutz,PRE2010Jeon,PRE2013Ralf,PRE2013Sylvia,JRheol2009Gregory}, where the system displays ballistic dynamics in the underdamped regime ($t\ll \tau_H$) and subdiffusion with an anomalous exponent of $\mu=2+2H$ in the overdamped regime ($t\gg \tau_H$).
On the other hand, the active component demonstrates multi-scale, complicated diffusion patterns that are strongly affected by the viscoelastic memory strength (quantified by $H$), as well as the values of $\tau_H$ and $\tau_\mathrm{A}$. When $1/2<H<3/4$, the active diffusion reveals power-law scalings, leading to two distinct MSD patterns depending on the relationship between $\tau_\mathrm{A}$ and $\tau_H$. 
In Case $1$ (where $\tau_\mathrm{A}>\tau_H$), the active component exhibits hyperdiffusion with $\mu=4$ in the underdamped regime, followed by the multiple overdamped motion, i.e., sub-ballistic active superdiffusion with $\mu=4-4H$ for $\tau_H\ll t\ll \tau_\mathrm{A}$ and active subdiffusion with $\mu=3-4H$ for $t\gg\tau_\mathrm{A}$ [Fig.~\ref{fgr3}(a)]. In Case $2$ (where $\tau_H>\tau_\mathrm{A}$), the underdamped dynamics became more complicated. It starts with hyperdiffusion with $\mu=4$ for $t\ll \tau_\mathrm{A}$, followed by another hyperdiffusion with $\mu=3$ for $\tau_\mathrm{A}\ll t\ll \tau_H$. In the overdamped limit, the active dynamics was simply subdiffusion with $\mu=3-4H$ [Fig.~\ref{fgr3}(a)]. 
In the case of $H>3/4$, the overdamped active dynamics deviate from power-law scalings. In both Cases $1$ \& $2$, the overdamped dynamics transformed into logarithmic ultraslow diffusion ($\sim \ln t$) at $H=3/4$ or confined diffusion for $3/4<H<1$. It is worth noting that the thermal dynamics dominate the active FLE systems in the limits of $t\to 0$ and $t\gg \tau^*$, while the active dynamics usually prevail in between the two limits. The cross-over times were sensitively determined by the strength of active noise ($D_\mathrm{A}$).

Notably, our analytic and numerical studies demonstrated that the active FLE system exhibits simpler dynamic characteristics when considering time-averaged observables. 
In the underdamped regime, the active component of TA MSD always displays ballistic diffusion for $t\ll\tau_H~(\&~\tau_\mathrm{A})$, regardless of the initial conditions and the values of $\tau_\mathrm{A}$ and $\tau_H$ [Fig.~\ref{fgr3}(c)]. This behavior is in stark contrast to the hyperdiffusions ($\langle x^2(t)\rangle\sim t^4$, $t^3$) observed in the EA MSDs. The ballistic dynamics then smoothly cross-overs to the overdamped dynamics, which are characterized by subdiffusion with a scaling of $t^{3-4H}$ for $1/2<H<3/4$, $\sim \ln t$ at $H=3/4$, and confined diffusion for $3/4<H<1$. Importantly, there is no sub-ballistic superdiffusion of $\sim t^{4-4H}$ observed in the overdamped counterpart of the active EA MSD when $\tau_\mathrm{A}>\tau_H$. Furthermore, in the overdamped regime for $t\gg\tau_H~\&~\tau_\mathrm{A}$, the TA MSD shares the same scaling behaviors with the EA counterpart. However, the amplitudes of both MSDs are not identical, indicating that the active component exhibits (ultra)weak ergodicity breaking (UWEB). We note that the observed UWEB in this work is distinguished from that reported in superdiffusive L{\'e}vy walk~\cite{PRE2013EBEli,PRL2013EBRalf}. In the latter, the UWEB arises due to insufficient self-averaging in finite measurements. Similarly, in the subdiffusive continuous-time random walk (CTRW) model, the lack of time scale in the waiting times results in WEB even at infinite measurement times~\cite{PRL2008Barkai,CTRW2013EuroL,PhToday2012CTRWEliRalf,Albers2013Eurolett}. In our active FLE system, the viscoelastic memory combined with a nonequilibrium noise plays a key role in inducing ergodicity-breaking.

We studied the ergodic properties of the active FLE systems using the ergodicity-breaking parameter, $\mathcal{EB}$, defined by Eq.~\eqref{eq:EB}. Our findings revealed that long-time viscoelastic memory plays a crucial role in determining the ergodic behaviors of these systems. (1) In the absence of viscoelastic memory (corresponding to the limit $H\to1/2$), the active LE--also known as the AOUP model~\cite{AOUP2022Lowen,PRE2019Bonilla}--exhibits transient ergodicity breaking at intermediate timescales in between $\tau_H$ and $\tau_\mathrm{A}$. Beyond this timescale, the viscoelasticity-free active system rapidly recovers ergodicity as time increases. 
(2) When the viscoelastic effect comes into play, the active FLE with $H\in(1/2,1)$ exhibits $H$-value sensitive ergodic dynamics. Despite the violation of ergodicity observed in the active dynamics [Fig.~\ref{fgr5}], the overall dynamics of the system, which combines the nonergodic active component with the thermal ergodic component, is always ergodic in the long run in the sense that $\mathcal{EB}\to1$ as $t$ increases beyond the cross-over time $\tau^*$ [Eq.~\eqref{eq:taustar}].
The observed ergodicity in the infinite-time regime simply reflects the ergodic property of the thermal component that dominates over the active component at large times of $t\gg \tau^*$.
However, before entering this ergodic regime, the active FLE system undergoes a nontrivial ergodicity-breaking state in the time window of $\tau_\mathrm{A} \lesssim t\lesssim \tau^*(H)$. Importantly, the magnitude of $\tau^*(H)$ is much larger than $\tau_\mathrm{A}$ even at moderate strengths of $D_\mathrm{A}$. As a result, the active FLE system is \textit{apparently} non-ergodic within a typical finite measurement time window (as observed in Figs.~\ref{fgr6}(b) \&~\ref{fgr6}(c)). Particularly in the domain of $3/4<H<1$, it becomes almost infeasible to observe the infinite-time ergodic regime. The cross-over time $\tau^*(H)$ diverges as $H\to 1$, as the analytic expressions~\eqref{tau*1} \&~\eqref{tau*2} reveal.
This divergence indicates that the apparent ergodicity-breaking state becomes exceptionally long-lived, making it challenging to observe ergodicity within practical measurement times. 

The apparent WEB in our active FLE system is a novel phenomenon that has not been reported in other systems. A well-known WEB is the discrepancy of the scaling relations between EA MSD ($\sim t^\alpha$) and TA MSD ($\sim t/T^{1-\alpha}$) in the CTRW or heterogeneous diffusion models~\cite{PRL2008Barkai,CTRW2013EuroL,PhToday2012CTRWEliRalf,Albers2013Eurolett,Cherstvy2013NJP,PRL2014ATTM}. In these models, the presence of long-lasting particle trapping events leads to non-stationary solution and breaks ergodicity.
Regarding the aforementioned UWEB in, e.g., the superdiffusive L{\'e}vy walk, EA and TA MSDs have the same scaling relation but different amplitudes. In contrast, our active FLE system displays a unique combination of these features. The active component exhibits UWEB, which is combined with the ergodic thermal component in the presence of a power-law viscoelastic memory. As a result, the violation of ergodicity persists during a remarkably long time window, with the duration expected to become infinite as $H$ approaches $1$.

The theories of the active FLE system developed in this work can be applied to quantitatively understand transport phenomena in various active viscoelastic systems. For example, it can be used to analyze the diffusion of AOUPs strongly bound (or cross-linked) to flexible polymers. In a computational investigation reported in Ref.~\cite{SoftMatter2020Joo}, it was found that, counterintuitively, the AOUPs interacting with flexible chains become more subdiffusive as their active mobility increases. Their analytic study demonstrated that the viscoelastic feedback from the flexible polymers induces an ultraslow subdiffusion with $\langle x_\mathrm{ac}^2(t)\rangle\sim \ln t$~\cite{SoftMatter2020Joo}. This type of AOUP systems can be described by our active FLE model with $H=3/4$ corresponding to the memory kernel for flexible polymers. In agreement with Ref.~\cite{SoftMatter2020Joo}, the active component of our active FLE system has EA and TA MSDs of $\sim \ln t$. Furthermore, our study predicts that the corresponding system exhibits significant ergodicity-breaking, characterized by $\mathcal{EB}_\mathrm{ac}=2$, indicating a visible discrepancy between the two MSDs in the intermediate timescale. However, the system recovers ergodicity at timescales shorter than those for other $H$ cases (see Fig.~\ref{fgr6}(d)). 

The diffusion of AOUPs connected to semiflexible filaments is evidently different from those in flexible polymers. Previous studies have reported that the overdamped diffusion dynamics of AOUPs in such a situation reveal scaling behaviors of $\langle x^2(t)\rangle\sim t^{3/2}$ for $t<\tau_\mathrm{A}$ and $\langle x^2(t)\rangle\sim t^{1/2}$ for $t>\tau_\mathrm{A}$~\cite{Han2023,Michael2000PRL}. The active FLE with $H=5/8$ (explaining the kernel for semiflexible filaments) with Case $1$ ($\tau_\mathrm{A}>\tau_H$) precisely explains these scaling behaviors [Eq.~\eqref{eq:xac2_case1}]. Our study gives further information about this system. For example, in the underdamped limit, the AOUP is expected to display ballistic movement ($\sim t^2$) as well as hyperdiffusion ($\sim t^4$). However, if its diffusion is observed with TA MSD, it will simply display the ballistic motion in the underdamped regime and a subdiffusion of $\sim t^{1/2}$ [Fig.~\ref{fgr3}]. The underdamped hyperdiffusion and the overdamped superdiffusion of $\sim t^{3/2}$ observed in the EA MSD will be masked. 

Our theoretical framework can be applicable to more complex active viscoelastic systems where multiple distinct viscoelastic memory effects are combined. For example, consider the diffusion of flexible polymers in an active viscoelastic bath, as reported in Refs.~\cite{PREVandebroek2015,Grimm2018Softmatter}. In these simulations, the center-of-mass (COM) of the polymer was observed to perform anomalous diffusion, with $\langle x_\mathrm{CM}^2(t)\rangle$ scaling as $\sim t^{4-4H}$ for $t<\tau_\mathrm{A}$ and $\sim t^{3-4H}$ for $t>\tau_\mathrm{A}$, where $\tau_\mathrm{A}$ is the persistence memory time of the active Ornstein-Uhlenbeck noise of the surrounding medium. The observed active diffusion of the COM can be understood as the overdamped diffusion of an AOUP in our active FLE model with Case $1$ ($\tau_\mathrm{A}>\tau_H$); see Eq.~\eqref{eq:xac2_case1}. 
Now consider the diffusion of a single monomer comprising the polymer. In this case, the dynamics of the single monomer are affected by two viscoelastic memory effects; one from the flexible polymer itself and the other from the surrounding viscoelastic medium. These two memory effects were found to induce complex diffusion dynamics of the single monomer, with $\langle x_\mathrm{m}^2(t)\rangle$ scaling as $\sim t^{\frac{3}{2}(2-2H)}$ for $t<\tau_\mathrm{A}$ and $\sim t^{\frac{3}{2}(2-2H)-1}$ for $t>\tau_\mathrm{A}$. The observed scaling relations can be effectively explained by considering the overdamped active diffusion of Case $1$ in our active FLE model, with the memory kernel modified as $K(t)\sim t^{\frac{3}{4}(2H-2)}$ to account for the combined effects of the two viscoelastic memory components.

Our active FLE model serves as a theoretical framework for quantitatively describing activity-induced transport dynamics in viscoelastic media, e.g., synthesized hydrogels, chromosomes, cytoskeletal networks, and the endoplasmic reticulum network. 
Expanding the current formalism to incorporate confining or random potentials could yield insights into various trapped-and-hopping phenomena within active biological systems, offering a fundamental understanding of the active viscoelastic transport in complex environments.

\section*{Acknowledgements}
This work was supported by the National Research Foundation
(NRF) of Korea, Grant No.~2021R1A6A1A10042944 \& No. RS-2023-00218927.

\appendix
\renewcommand{\thefigure}{A\arabic{figure}}
\setcounter{figure}{0}
\setcounter{equation}{0}
\begin{figure*}
      
       \includegraphics[width=0.8\textwidth]{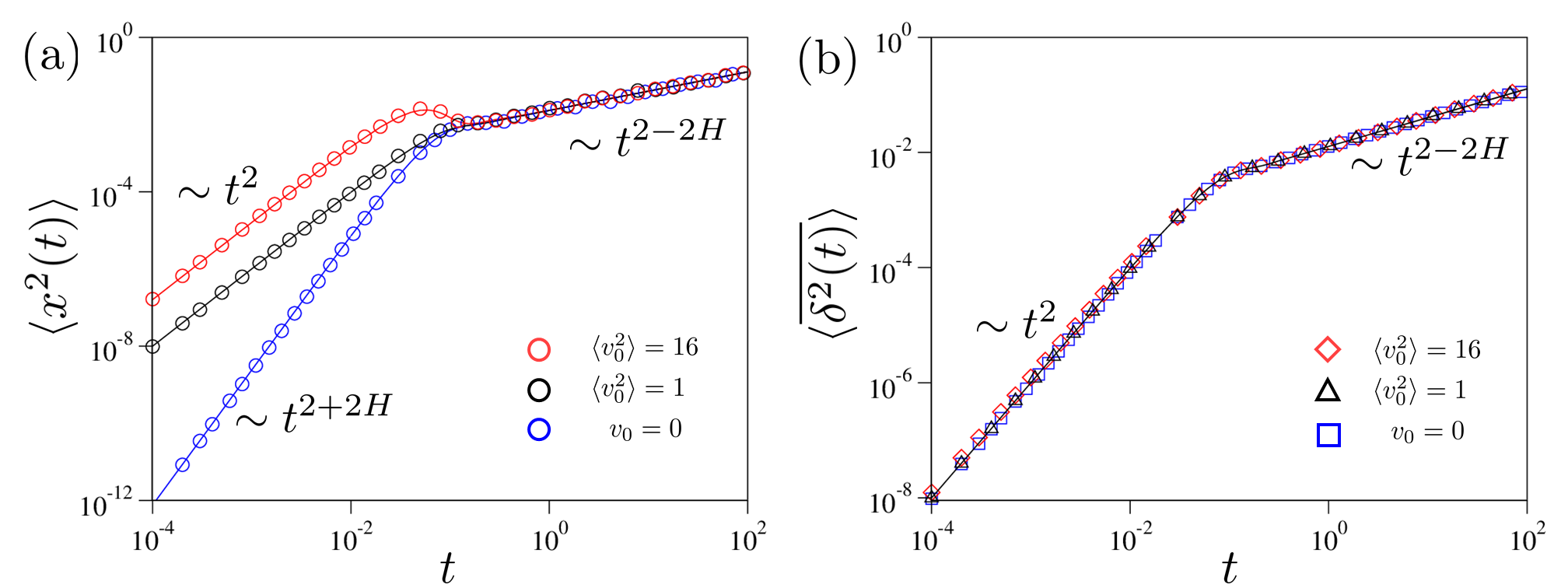}
      
     \caption{The ensemble- and time-averaged MSDs for the equilibrium FLE model with the three initial conditions: $v_0=0$, $\langle v_0^2\rangle=\langle v^2\rangle_\mathrm{th}=1$ (the equilibrium initial condition), and $\langle v_0^2\rangle=16$. In the two panels, $\tau_H=0.03$ with $H=3/4$. (a) The simulation results (symbols) for $\langle x^2(t)\rangle$ are plotted together with the theory~\eqref{eq:msdFDTv0} (solid lines). When $t<\tau_H$, the EA MSDs show ballistic dynamics with $v_0\neq0$ and the hyperdiffusion with the anomalous exponent of $2+2H(>2)$ at $v_0=0$. For $t>\tau_H$, all the MSDs exhibit subdiffusion, perfectly agreeing with Eq.~\eqref{eq:msdFDT}. (b) The simulation results for $\langle \overline{\delta^2(t)}\rangle$s (symbols). The three data follow the solid line representing Eq.~\eqref{eq:delta2_FDT}, regardless of the initial velocities. }
     \label{figA1}
\end{figure*}
\section{Validation of numerical method \&\\ FLE with thermal noise}\label{sec:appendixA} 
Let us introduce a numerical algorithm to solve the active FLE~\eqref{eq:FLEwoFDT2}, using the method introduced in~\cite{PRE2009Eli,PRE2010Jeon,Numerics2013Guo}. Firstly, by integrating Eq.~\eqref{eq:FLEwoFDT2} from $0$ to $t$, we obtain the stochastic integral equation for velocity $v(t)=\dot{x}(t)$ with the thermal noise $\eta\xi_H(t)$ and active noise $\eta_\mathrm{A}\xi_\mathrm{A}(t)$, given by
\begin{equation}\label{eq:SIE}
\begin{aligned}
   v(t)=&v_0-\frac{\gamma_0}{m}\int_0^t\frac{(t-t')^{2H-1}}{2H-1}v(t')dt'\\&+\frac{\eta}{m}\int_0^t \xi_H(t')dt'+\frac{\eta_\mathrm{A}}{m}\int_0^t \xi_\mathrm{A}(t')dt'.  
\end{aligned}
\end{equation} 
To numerically calculate the stochastic integral equation~\eqref{eq:SIE}, we employ the product trapezoidal quadrature formula on discrete time points $t_n$ ($n=0,~1,~\dots,~N$) within the time domain $[0,~T]$~\cite{Diethelm2002}:

\begin{equation}\label{eq:vdiscrete}
\begin{aligned}
    v(t_{n+1})=&v_0-\frac{\gamma_0}{m}\frac{h^{2H}}{(2H+1)!}\sum_{j=0}^{n+1}a_{j,n+1}v(t_j)\\&+\frac{\eta}{m}\sum_{j=0}^{n} \xi_H(t_j)+\frac{\eta_\mathrm{A}}{m}\sum_{j=0}^{n} \xi_\mathrm{A}(t_j).
\end{aligned}
\end{equation}
The Eq.~\eqref{eq:vdiscrete} provides the discrete form of the velocity, where $h(=t_{n+1}-t_n)$ represents the time increment, and the coefficients $a_{j,n+1}$ are $n^{2H+1}-(n-2H)(n+1)^{2H}$ for $j=0$, $(n-j+2)^{2H+1}+(n-j)^{2H-1}-2(n-j+1)^{2H+1}$ for $1\le j\le n$, and $1$ for $j=n+1$.
Rearranging Eq.~\eqref{eq:vdiscrete}, we find the expression for the velocity at $t_{n+1}$ in the following:
\begin{equation}\label{eq:vcomputation}  
    \begin{aligned}
        v(t_{n+1})=&\left(\frac{m(2H+1)!}{m(2H+1)!+h^{2H}\gamma_0}\right)\\      &\times\left(v_0-\frac{\gamma_0}{m}\frac{h^{2H}}{(2H+1)!}\sum_{j=0}^{n}a_{j,n+1}v(t_j)\right.\\
        &\qquad+\left.\frac{\eta}{m}\sum_{j=0}^{n}\xi_H(t_j)+\frac{\eta_\mathrm{A}}{m}\sum_{j=0}^{n}\xi_\mathrm{A}(t_j)\right).
    \end{aligned}
\end{equation}

Once $v(t_{n+1})$ is obtained, the trajectory on $t_{n+1}$ is evaluated using the second-order Euler method
\begin{equation}\label{eq:xcomputation}
   x(t_{n+1})=x(t_n)+\frac{h}{2}(v(t_n)+v(t_{n+1})).
\end{equation}

In our active FLE model, we have two noise terms, i.e., $\eta\xi_{H}(t)$ and $\eta_\mathrm{A}\xi_\mathrm{A}(t)$. The former noise is fGn explained in Sec.~\ref{sec2}, and the summation of fGn, i.e., $\sum_{j=0}^{n}\xi_H(t_j)$ in Eq.~\eqref{eq:vcomputation}, produces fractional Brownian motion (fBm) $x_{H}(t)\equiv\int_0^t\xi_{H}(t')dt'$, which has zero mean of $\langle x_{H}(t) \rangle=0$ and the covariance $\langle x_{H}(t_1)x_{H}(t_2) \rangle=D_{H}\left(|t_1|^{2H}+|t_2|^{2H}-|t_1-t_2|^{2H}\right)$. In our simulation, fBm is generated using the \textsf{FFGN} function implemented in Matlab~\cite{FFGN1987,FFGN1998,FFGN2002Bardet,FFGN2007} for a given trajectory length $N$, the Hurst exponent $H$, and $D_{H}$. In our simulation, we set $D_{H}=1/2$. Note that the simulation of the active FLE does not depend on the value of $D_{H}$ because the prefactor $\eta\propto \sqrt{1/D_{H}}$ cancels out the $D_{H}$-dependency. The active OU noise $\xi_\mathrm{A}(t_n)$ in Eq.~\eqref{eq:vcomputation} is obtained by solving the stochastic equation~\eqref{eq:genaoup2} in a discrete version, which is
\begin{equation}\label{eq:genaoup3}
    \xi_\mathrm{A}(t_{n+1})=\left(1-\frac{h}{\tau_\mathrm{A}}\right)\xi_\mathrm{A}(t_n)+\sqrt{\frac{2hD_\mathrm{A}}{\tau_\mathrm{A}}}\omega
\end{equation}
where $\omega$ is a random number chosen from a normal distribution $\mathcal{N}(0,1)$.

We validate our code by simulating the equilibrium FLE model (neglecting the active noise part by setting $D_\mathrm{A}=0$ in the active FLE~\eqref{eq:FLEwoFDT2}) with various initial conditions. We measure the ensemble-averaged MSD and compare it with theoretical expectations. The MSD is defined as $\langle x^2(t)\rangle\equiv\int_0^tdt_1\int_0^tdt_2\langle v(t_1)v(t_2)\rangle$, and for a stationary process we have $\langle x^2(t)\rangle=2\int_0^tdt_1\int_0^{t_1}d\tau\langle v(\tau)v(0)\rangle$
with $\tau=|t_1-t_2|$. The Laplace transform of the MSD follows a simple relation 
\begin{equation}\label{eq:msdvac}
     \mathcal{L}[\langle x^2(t)\rangle](s)=\frac{2}{s^2}\mathcal{L}[\langle v(t)v(0)\rangle](s)=\frac{2\langle v_0^2 \rangle}{s^3+\tau_H^{-2H} s^{3-2H}},
\end{equation} 
and its inverse transform leads to the the exact solution expressed in terms of a generalized Mittag-Leffler function
 \begin{equation}\label{eq:msdFDT}
     \langle x^2(t)\rangle=2\langle v^2\rangle_\mathrm{th} t^2E_{2H,3}\left[-\left(\frac{t}{\tau_H}\right)^{2H}\right].
 \end{equation}
By the definition of the generalized Mittag-Leffler function~\eqref{eq:MLF}, we have $\langle x^2(t)\rangle\approx\langle v_0^2\rangle t^2$ for the short-time limit. The long-time asymptotic expression for the MSD is found to be $\langle x^2(t)\rangle\approx2v_0^2t^{2-2H}/\Gamma(3-2H)$ from the series expansion~\eqref{eq:MLFlong}. Thus, the MSD shows a cross-over from the ballistic motion at short times to the subdiffusive motion at long times, which has the form~\cite{PRE2009Eli,PRE2010Jeon}
\begin{equation}\label{eq:msdlimit}
\langle x^2(t)\rangle\approx\frac{2k_B\mathcal{T}}{m}\left\{\begin{array}{ll}
      \frac{1}{\Gamma(3)}t^2,& t \to0 \\
      \frac{\tau_H^{2H}}{\Gamma(3-2H)}t^{2-2H},&t \to\infty. 
\end{array}\right.     
\end{equation}

The exact analytic expression of the MSD is reported in some previous work~\cite{PRE2013Ralf,Bao2017}. When the initial condition is not equilibrium, i.e., $\langle v_0^2\rangle \neq k_B\mathcal{T}/m$, the MSD is written as 
\begin{equation}\label{eq:msdFDTv}   
\begin{aligned}
    \langle x^2(t)\rangle=&2\langle v^2\rangle_\mathrm{th} t^2E_{2H,3}\left[-\left(\frac{t}{\tau_H}\right)^{2H}\right]\\
    &+\left(\langle v_0^2\rangle-\langle v^2\rangle_\mathrm{th} \right)t^2E_{2H,2}\left[-\left(\frac{t}{\tau_H}\right)^{2H}\right]^2
\end{aligned}
\end{equation}
with $\tau=|t-t'|$.
If $\langle v_0^2\rangle=\langle v^2\rangle_\mathrm{th}$ or $t\to\infty$, Eq.~\eqref{eq:msdFDTv} yields Eq.~\eqref{eq:msdFDT}.
For short times $t\ll\tau_H$, $\langle x^2(t)\rangle\sim\langle v_0^2\rangle t^2$ for $v_0\neq0$, while, for $v_0=0$, hyperdiffusion is expected with $\langle x^2(t)\rangle\sim t^{2+2H}$. 

For both equilibrium and nonequilibrium initial conditions, the time-averaged MSD is obtained to be~\cite{PRE2009Eli,PRE2010Jeon} 
\begin{equation}\label{eq:delta2_FDT}
\langle \overline{\delta^2(t)}\rangle=2\langle v^2\rangle_\mathrm{th} t^2E_{2H,3}\left[-\left(\frac{t}{\tau_H}\right)^{2H}\right],
\end{equation} 
which is the same functional form as Eq.~\eqref{eq:msdFDT}. In Fig.~\ref{figA1}, we simulate the thermal FLE systems with several different initial conditions and compare them with the exact theoretical expressions above. As shown, the simulation data show excellent agreement with the theory for all the three cases.

 \section{Generalized Mittag-Leffler functions}\label{sec:appendixB}

Here we introduce the generalized Mittag-Leffler function (MLF), $E_{a,b}(z)$, and some important properties of this function. The generalized MLF is defined by the series expansion as follows~\cite{MLFunction1955,MLfunction2011,Jia2019}:
\begin{equation}\label{eq:MLFApendix}
    E_{a,b}(z)=\sum_{k=0}^\infty\frac{z^k}{\Gamma(a k+b)}
\end{equation} 
where $a,b\in \mathbb{C}$, $\mathcal{R}(a)>0$, $\mathcal{R}(b)>0$, $z\in \mathbb{C}$ with $\mathbb{C}$ being the set of complex numbers.
For large-$z$ expansion, the generalized MLF is expressed as 
\begin{equation}\label{eq:MLFlong}
     E_{a,b}(z)=-\sum_{k=1}^\infty\frac{z^{-k}}{\Gamma(b-a k)}.
\end{equation} 
 When $a=1$ and $b=1$, the MLF reduces to the exponential function $E_{1,1}(z)=e^z$. Using the above series expansion formulae, we find that $E_{a,b}[-ct^{a}]\approx \exp[-ct^a/\Gamma(a+b)]$ for $t\ll1$ and $\approx t^{-a}/\Gamma(b-a)$ for $t\gg1$. 

For the Mittag-Leffler function $E_{a,b}[-ct^{a}]$, a useful Laplace transform formula is the following:
 \begin{equation}
    \mathcal{L}\left[t^{b-1}E_{a,b}(\pm c t^a)\right](s)=\frac{s^{-b}}{\left(1\mp c s^{-a}\right)}.
\end{equation}
We employ this formula to compute the Laplace transform of the convolution forms (such as Eqs.~\eqref{eq:v_th},~\eqref{eq:v_ac},~\eqref{eq:xeq} \&~\eqref{eq:xac}) that involve the generalized MLF.
Another useful property of the MLF is its relation to the derivation of MLF:
\begin{equation}
    \frac{d^n}{dt^n}\left[t^{b-1}E_{a,b}(c t^a)\right]=t^{b-n-1}E_{a,b-n}(c t^a)
\end{equation}
where $b\ge n+1$. 

\begin{figure*}
\centering
\includegraphics[width=1.0\textwidth]{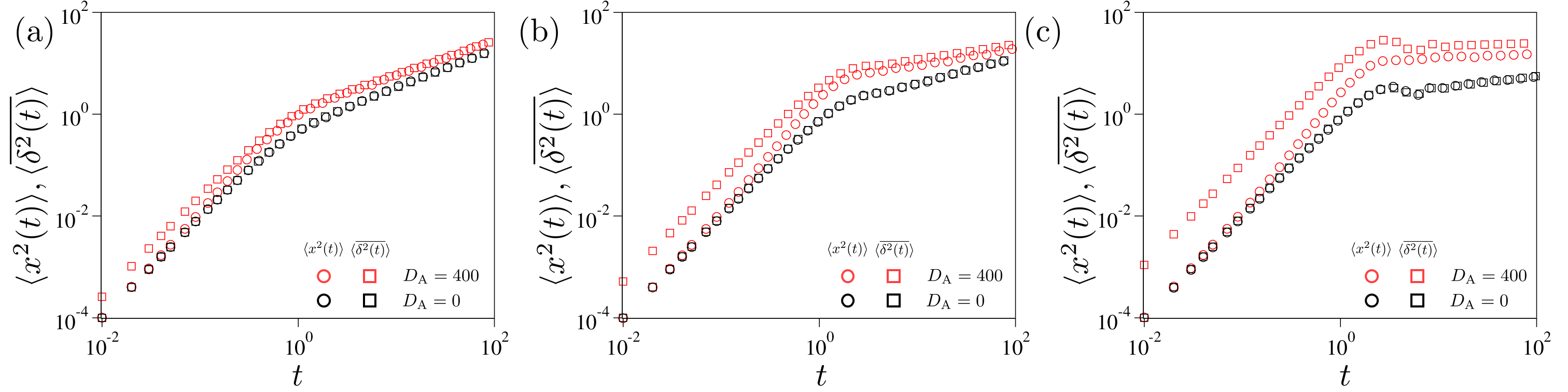}
\caption{The comparison of $\langle x^2(t)\rangle$ and $\langle\overline{\delta^2(t)}\rangle$ for Case $2$ ($\tau_\mathrm{A}\ll\tau_H$). In each plot, the simulation results for $D_\mathrm{A}=0$ and $400$ are present. (a) $H=5/8$ with $\tau_\mathrm{A}=0.01$ and $\tau_H=0.36$. (b) $H=3/4$ with $\tau_\mathrm{A}=0.01$ and $\tau_H=0.68$. (c) $H=7/8$ with $\tau_\mathrm{A}=0.01$ and $\tau_H=0.88$. For all panels, the simulation is performed with an initial condition of $\langle v_0^2\rangle=1$.}
\label{figureC1}
\end{figure*}

\begin{figure}
    \centering
    \includegraphics[width=0.45\textwidth]{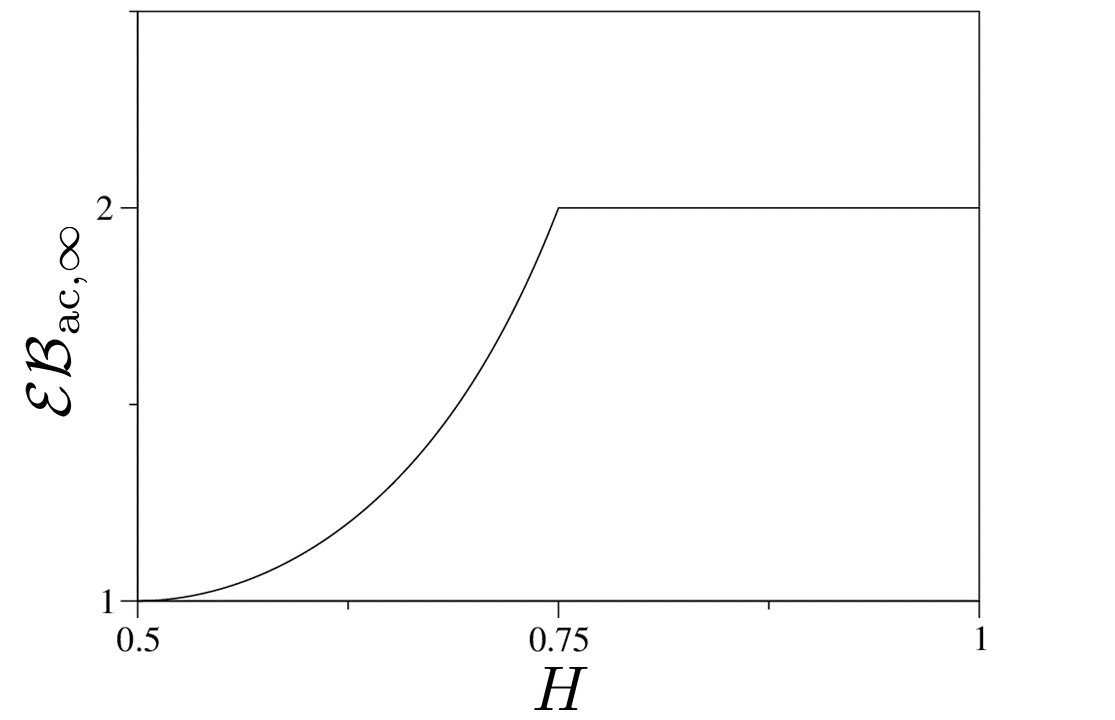}
    \caption{$\mathcal{EB}_{\mathrm{ac},\infty}$, Eq.~\eqref{eq:EB_ac_inf}, is plotted as a function of $H$. $\mathcal{EB}_{\mathrm{ac},\infty}$ monotonically increases as $H$ increases in the range of $1/2<H<3/4$. For $H\geq 3/4$, $\mathcal{EB}_{\mathrm{ac},\infty}=2$ }
    \label{figureC2}
\end{figure}

\section{Supplementary figures}

In this section, we plot supplementary figures supporting our main results.
In Fig.~\ref{figureC1}, we present the ensemble- and time-averaged MSDs for Case $2$ ($\tau_H>\tau_\mathrm{A}$) with the Hurst exponent $H=5/8$ (a), $3/4$ (b), and $7/8$ (c). In three panels, the EA and TA MSDs are simulated with $D_\mathrm{A}=0$ and $400$ with a non-stationary initial condition of $\langle v_0^2\rangle=1$. In the thermal FLE process ($D_\mathrm{A}=0$), both MSDs are identical. For the active FLE process ($D_\mathrm{A}=400$), the EA and TA MSDs deviate from each other, and the discrepancies increase as $H$ increases. Note that TA MSDs are always greater tan the EA MSDs across all times.

In Fig.~\ref{figureC2}, we plot $\mathcal{EB}_{\mathrm{ac},\infty}$ as a function of $H$ (see Eq.~\eqref{eq:EB_ac_inf}). At 
$H=1/2$, the system becomes the conventional AOUP model in viscous media, which is ergodic, thus, resulting in $\mathcal{EB}_{\mathrm{ac},\infty}=1$. As $H$ is in the range $1/2<H<3/4$, we observe a monotonic increase in $\mathcal{EB}_{\mathrm{ac},\infty}$ with increasing $H$, eventually reaching a value of $2$ as $H\to3/4$. At $H=3/4$, $\mathcal{EB}_{\mathrm{ac},\infty}=2$ originates from the logarithmic growths of both MSDs (Eqs.~\eqref{eq:xac2_long} \&~\eqref{eq:tamsdlong2}). If $H>3/4$, the confined motion of MSD maintains $\mathcal{EB}_{\mathrm{ac},\infty}=2$ (Eq.~\eqref{eq:EB_ac_inf}).


\section*{References}
%
\bibliography{GLE}
\end{document}